\documentclass[longauth]{aa}

\usepackage{txfonts}
\usepackage{pgf}
\usepackage{graphicx}
\usepackage{capt-of}
\usepackage{longtable}
\usepackage{multirow}
\let\linenumbers\relax
\usepackage{xcolor}
\usepackage[title,toc]{appendix}
\usepackage{CJK}
\usepackage{natbib,twoopt}
\usepackage[breaklinks=true]{hyperref} %
\usepackage{amsmath}
\usepackage[utf8]{inputenc}
\makeatletter
\renewcommand{\aa@subabstractvname}{}  
\renewcommand{\aa@abstract@v}{\relax}  
\makeatother
\begin{document}

\title{Optical and Near-infrared Observations of SN\,2023ixf \\for over 600 days after the Explosion
}

\author{Gaici Li \inst{\ref{inst1}}
\and Xiaofeng Wang \inst{\ref{inst1}\thanks{E-mail: wang\_xf@mail.tsinghua.edu.cn}} 
\and Yi Yang \inst{\ref{inst1}, \ref{Berkeley} } 
\and A. Pastorello \inst{\ref{Reguitti2}} 
\and A. Reguitti \inst{\ref{Reguitti1}, \ref{Reguitti2}} 
\and G. Valerin \inst{\ref{Reguitti2}} 
\and P. Ochner \inst{\ref{DFA UNIPD}, \ref{Reguitti2}} 
\and Yongzhi Cai  \inst{\ref{instYNO},\ref{CentreSN},\ref{KeyObjects}} 
\and T. Iijima \inst{\ref{Reguitti2}} 
\and U. Munari \inst{\ref{Reguitti2}} 
\and I. Salmaso \inst{\ref{Salmaso_1},\ref{Reguitti2}} 
\and A. Farina \inst{\ref{DFA UNIPD}} 
\and R. Cazzola \inst{\ref{DFA UNIPD}} 
\and N. Trabacchin \inst{\ref{Trabacchin}} 
\and S. Fiscale \inst{\ref{Fiscale1}, \ref{Fiscale2}} 
\and S. Ciroi \inst{\ref{DFA UNIPD},\ref{Reguitti2}} 
\and A. Mura \inst{\ref{DFA UNIPD},\ref{Reguitti2}} 
\and A. Siviero \inst{\ref{DFA UNIPD},\ref{Reguitti2}} 
\and F. Cabras \inst{\ref{DFA UNIPD}} 
\and M. Pabst \inst{\ref{DFA UNIPD}} 
\and S. Taubenberger \inst{\ref{MPA}} 
\and C. Vogl \inst{\ref{MPA}} 
\and C. Fiorin 
\and Jialian Liu \inst{\ref{inst1}} 
\and Liyang Chen \inst{\ref{inst1}} 
\and Danfeng Xiang \inst{\ref{Beijing Planetarium}, \ref{inst1}} 
\and Jun Mo \inst{\ref{inst1}} 
\and Liping Li  \inst{\ref{instYNO},\ref{CentreSN},\ref{KeyObjects}} 
\and Zhenyu Wang \inst{\ref{instYNO},\ref{CentreSN},\ref{KeyObjects},\ref{UCAS}} 
\and Jujia Zhang \inst{\ref{instYNO},\ref{CentreSN},\ref{KeyObjects}} 
\and Qian Zhai \inst{\ref{instYNO},\ref{KeyObjects}} 
\and D.O. Mirzaqulov \inst{\ref{AZT1}} 
\and S.A. Ehgamberdiev  \inst{\ref{AZT1}, \ref{AZT2}} 
\and Alexei V. Filippenko \inst{\ref{Berkeley}}
\and Shengyu Yan \inst{\ref{inst1}} 
\and Maokai Hu \inst{\ref{inst1}}
\and Xiaoran Ma \inst{\ref{inst1}} 
\and Qiqi Xia \inst{\ref{inst1},\ref{Shandong}} 
\and Xing Gao \inst{\ref{Xingming}} 
\and Wenxiong Li \inst{\ref{Guotai}} 
}
\institute{
Department of Physics, Tsinghua University, Beijing, 100084, China. \label{inst1}
\and Department of Astronomy, University of California, Berkeley, CA 94720-3411, USA \label{Berkeley}
\and INAF - Osservatorio Astronomico di Padova, Vicolo dell'Osservatorio 5, 35122 Padova, Italy \label{Reguitti2}
\and INAF - Osservatorio Astronomico di Brera, Via E. Bianchi 46, 23807 Merate (LC), Italy \label{Reguitti1}
\and Dipartimento di Fisica e Astronomia, Universit\`a Degli Studi di Padova, 35121 Padova PD, Italy\label{DFA UNIPD}
\and Yunnan Observatories, Chinese Academy of Sciences, Kunming 650216, China \label{instYNO}
\and International Centre of Supernovae, Yunnan Key Laboratory, Kunming 650216, China \label{CentreSN}
\and Key Laboratory for the Structure and Evolution of Celestial Objects, Chinese Academy of Sciences, Kunming 650216, China \label{KeyObjects}
\and INAF-Osservatorio Astronomico di Capodimonte, Salita Moiariello 16, 80131 Napoli, Italy \label{Salmaso_1}
\and CISAS G. Colombo, University of Padova, via Venezia 15, 35131, Padova Italy \label{Trabacchin}
\and UNESCO Chair ''Environment, Resources and Sustainable Development'', Department of Science and Technology, Parthenope University of Naples, Italy \label{Fiscale1}
\and INAF, Osservatorio Astronomico di Capodimonte, Salita Moiariello, 16, Naples, I-80131, Italy \label{Fiscale2}
\and Max-Planck-Institut fur Astrophysik, Karl-Schwarzschild Straße 1, 85748 Garching, Germany \label{MPA}
\and Beijing Planetarium, Beijing Academy of Sciences and Technology, Beijing 100044, China \label{Beijing Planetarium}
\and University of Chinese Academy of Sciences, Beijing 100049, China \label{UCAS}
\and Ulugh Beg Astronomical Institute, Astronomy Street 33, Tashkent 100052, Uzbekistan \label{AZT1}
\and National University of Uzbekistan, Tashkent 100174, Uzbekistan \label{AZT2}
\and School of Space Science and Technology, Shandong
University, Weihai, Shandong 264209, China \label{Shandong}
\and Xingming Observatory, Urumqi, 830000, China \label{Xingming}
\and National Astronomical Observatories, Chinese Academy of Sciences, Beijing 100101, People's Republic of China \label{Guotai}
}
\abstract
{We present a comprehensive photometric and spectroscopic study of the nearby Type II supernova (SN) 2023ixf, with our extensive observations spanning the phases from $\sim$3 to over 600 days after the first light.}
{The aim of this study is to obtain key information on the explosion properties of SN\,2023ixf and the nature of its progenitor.}
{The observational properties of SN\,2023ixf are compared with those of a representative sample of Type IIP/IIL SNe to investigate commonalities and diversities. We conduct a detailed analysis of temporal evolution of major spectral features observed throughout different phases of the SN\,2023ixf explosion. Several key interpretations are addressed through a comparison between the data and the model spectra predicted by non-local thermodynamic equilibrium (non-LTE) radiative-transfer calculations for progenitor stars within a range of zero-age main sequence (ZAMS) masses.}
{Our observations indicate that SN\,2023ixf is a transitional SN that bridges the gap between Type IIP and IIL subclasses of H-rich SNe, characterized by a relatively short plateau ($\lesssim 70$\,d) in the light curve. 
It shows a rather prompt spectroscopic evolution toward the nebular phase; emission lines of Na, O, H, and Ca in nebular spectra all exhibit multipeak profiles, which might be attributed to bipolar distribution of the ejecta. In particular, the H$\alpha$ profile can be separated into two central peaked components (with a velocity of about 1500\,km\,s$^{-1}$) that is likely due to nickel-powered ejecta and two outer peak/box components (with a velocity extending up to $\sim 8000$\,km\,s$^{-1}$) that can arise from interaction of the outermost ejecta with a circumstellar shell at a distance of $\sim 6.2 \times 10^{15}$\,cm.  
The nebular-phase spectra of SN\,2023ixf show good agreement with those predicted by a non-LTE radiative-transfer code for progenitor stars with a ZAMS mass ranging from 15 to 19\,M${_\odot}$. A distance $D = 6.35^{+0.31}_{-0.39}$\,Mpc is estimated for M101 based on the expanding photosphere method.}


\maketitle

\section{Introduction}~\label{sec:intro}
Core-collapse supernovae (CCSNe) are terminal explosions of massive stars with zero-age main sequence (ZAMS) masses larger than $\sim 8$\,M$_{\odot}$. 
Observational properties among H-rich (or Type II) CCSNe span a broad parameter space, suggesting intrinsic diversity in progenitor properties and explosion mechanisms~\citep{2009MNRAS.395.1409S, 2009ARA&A..47...63S, 2011MNRAS.412.1441L, 2015PASA...32...16S}.
Two subclasses have been established solely based on the light-curve morphology, namely Type II-linear (SN IIL) and Type II-plateau (SN IIP; \citealp{1979A&A....72..287B}). 
The former shows a post-peak linear magnitude decline at a pace of $\sim$0.5 mag\,100\,d$^{-1}$, lasting $\lesssim$80--100 days (e.g., ~\citealp{2011MNRAS.412.1441L, Anderson2014}), while the latter maintains a rather constant brightness for $\sim$100--140 days after the light-curve peak \citep{Anderson2014}. Such a light-curve plateau can be understood as a consequence of the emission released by recombination of hydrogen in the outermost envelope of the SN ejecta~\citep{1993ApJ...414..712P, 2003MNRAS.338..711Z, 2010ApJ...725..904N}.
These modeling efforts suggest that the difference in the light-curve morphology of the SN IIP and SN IIL subclasses can be attributed to the remaining mass and radial density profile of the H-rich envelope of their massive progenitors.

It also remains debated whether the light-curve decline rates of SNe IIP and IIL follow a distinct \citep{2012ApJ...756L..30A, 2014MNRAS.445..554F} or continuous \citep{Anderson2014, 2015ApJ...799..208S} distribution.
In some cases, spectra of SNe II present a series of strong hydrogen Balmer lines with a low velocity ($\sim$20--800\,km\,s$^{-1}$; see \citealp{Smith_handbook,2023MNRAS.519.1618Y}, and references therein), which can 
remain visible even years after the explosion. These objects have been classified as Type II-narrow (IIn) SNe~\citep{Schlegel_1990},
for which the narrow Balmer lines arise from the
ionized circumstellar matter (CSM) previously expelled from the progenitor star. Such an enrichment procedure of the CSM may exhibit a continuous wind-like process or multiple episodes of mass ejections prior to the terminal explosion.

Recent observations of SNe II starting $\lesssim$ 1 day after the explosion have identified elevated mass loss immediately before the explosion. 
As a result, the ambient CSM will manifest as shock-breakout flash-ionized spectral features \citep{2007ApJ...666.1093Q, Gal-Yam_2013cu, 2017NatPh..13..510Y, 2024Natur.627..759Z}, which have been found in more than half of their massive progenitors \citep{2021ApJ...912...46B, 2023ApJ...952..119B}. 
As the ionization states of various atoms in the CSM are highly dependent on the injected energy, and they can be quickly swept away by the rapidly expanding ejecta owing to their vicinity, these ``flash'' features may persist for only a few days. 
Therefore, the line species identified in the prompt, flash-ionized spectral features closely trace the surface chemical composition of the exploding progenitor. The temporal evolution of the line strength also results from an interplay between the pre-explosion mass-loss history and the radial profile of 
the ionization states in the CSM.
Despite its ubiquity, however, the mechanism driving such enhanced mass loss within years of the terminal explosion remains unclear.
When the ejecta engulf the CSM, the outermost layer is dominated by the expanding H envelope, displaying broad P~Cygni profiles of Balmer lines.

Apart from the enigmatic enhanced pre-explosion mass loss, efforts have also been made to investigate the core-collapse process. 
Recent three-dimensional (3D) numerical calculations have been successful in exploding massive stars. After the initial core collapse, the outward transport of energy produced by the initial bounce shock is facilitated by absorbing neutrinos~\citep{Burrows_2021, 2024ApJ...964L..16B, 2024ApJ...969...74W, 2024arXiv241103434V}. 
Alternative models may involve energy deposition in the stellar envelope and the launch of moderately relativistic jets~\citep{1970ApJ...161..541L, Khokhlov_1999, 2003ApJ...598.1163M, 2014MNRAS.445.3169O, 2015Natur.528..376M, 2016ARNPS..66..341J, 2016NewAR..75....1S}.

While spectrophotometric monitoring starting from the earliest phases provides critical characteristics of the energy, chemical content, and kinematics of the SN explosion, observations at later phases yield highly complementary information that is essential to probe the explosion physics. 
After most of the H envelope has recombined, the late-time emission is powered by the radioactive decay of iron-group elements such as Ni and Co. 
These heavy nuclei were synthesized by the shock near the explosion center. 
As the nucleosynthesis depends strongly on the ZAMS mass of the star (e.g., \citealp{1995ApJ...448..315W}), the spectral features of the newly synthesized elements (e.g., C, O, Si, Ca, Fe, Co, and Ni) provide unique fingerprints of the explosion physics at work. 
Therefore, the line species, the ionization states, and the line profiles may provide critical clues about (respectively) the explosion core's chemical composition, physical conditions, and dynamics, thereby linking the progenitor properties and the explosion mechanism (e.g., \citealp{2017hsn..book..795J}). 
The deep interior is revealed only as the optical depth of the ejecta progressively decreases with time.

SN\,2023ixf was discovered as a young SN candidate in the nearby spiral galaxy Messier 101 (M101, also known as NGC\,5457) on 2023-05-19 17:27:15 (UTC dates are used throughout this paper) / MJD 60083.72726 by Koichi Itagaki~\citep{2023TNSTR1158....1I}. In this paper, we adopt a distance to M101 of $6.85 \pm 0.15$\,Mpc \citep{Riess_distance}. 
The first classification spectrum obtained at 2023 05-19 22:23 / MJD 60083.93 shows a set of narrow emission lines of H, He, C, and N superimposed on a
blue continuum~\citep{2023TNSCR1164....1P}. 
The profiles of these emission lines have full-widths at half-maximum (FWHMs) of $\lesssim 1000$--2000\,km\,s$^{-1}$ and centered at zero velocity relative to the redshift $z= 0.000804$ of its host~\citep{Riess_distance}. 
The observational signatures suggested a young, Type II nature for SN\,2023ixf~\citep{2023TNSCR1164....1P}.
The narrow features persisted for the first few days, with the most prominent H$\alpha$ line being still visible $\sim$10 days after the explosion. Such signatures in the early spectroscopic time sequence of SN\,2023ixf suggest that the progenitor experienced an enhanced mass loss prior to the explosion, building up dense CSM that extends up to  $\sim 10^{14}$\,cm~\citep{Bostroem_23ixf, Daichi_23ixf, Smith_23ixf, Teja_23ixf, Zhang_23ixf, Zheng_23ixf}.

The progenitor of SN\,2023ixf has been identified in archival images obtained by the {\it Hubble Space Telescope (HST)} and the {\it Spitzer Space Telescope (Spitzer)}. 
Comparisons between the fitting to the spectral energy distribution (SED) of the pre-explosion source and stellar evolutionary tracks indicate a red supergiant (RSG) progenitor enshrouded by a dusty shell. However, a broad mass range of the progenitor has been found, $\sim$8 to 20\,M$_{\odot}$~\citep{Pledger_progenitor, Kilpatrick_progenitor, Xiang_progenitor, Van_Dyk_progenitor, Jencson_progenitor, Niu_progenitor, Qin_progenitor, Soraisam_progenitor}. 

Several unprecedented datasets also provide key diagnostics of the explosion geometry. 
For example, a series of high-resolution spectra within the first week of the explosion revealed discrepancies among the evolution of the spectral profiles measured from different ionization states, which can be attributed to asymmetry in the dense CSM~\citep{Smith_23ixf}. 
The dramatically changing spectropolarimetric properties of SN\,2023ixf measured between $\sim$1.4 and 14.5 days after the explosion depict aspherical, optically thick CSM swept away by the aspherical SN ejecta within the first $\sim 5$ days~\citep{Vasylyev_2023ixf}. 
The initial reddish color that evolved blueward measured from as early as $\sim$1.4\,hr after the explosion is indicative of the gradual sublimation of pre-SN dust grains as the shock-breakout front propagates through an aspherical shell of CSM~\citep{2024Natur.627..754L}. 

The comparisons of the nebular spectra of SN\,2023ixf with model spectra suggest that its progenitor mass ranges from 10 to 16\,M$_{\odot}$~\citep{Michel_2023ixf, Kumar_2023ixf, Folatelli_2023ixf, Zheng_23ixf, Fang_ixf, Ferrari_2023ixf}. A boxy-shaped emission of H$\alpha$ is observed during the nebular phase, which hints at interaction between the ejecta and a CSM shell~\citep{Michel_2023ixf, Kumar_2023ixf, Folatelli_2023ixf, Zheng_23ixf, Ferrari_2023ixf}.

This paper presents extensive optical photometry and optical/near-infrared (NIR) spectroscopy of SN\,2023ixf spanning from the first day to over 600\,days after the explosion.
Our observations and data reduction are outlined in Section~\ref{sec:obs}. 
In Section~\ref{sec:lc} we discuss the photometric evolution. 
Section~\ref{sec:spec} details the spectral evolution. 
Implications of the observational properties are discussed in Section~\ref{sec:discussion}, and our concluding remarks are provided in Section~\ref{sec:conclusion}.

\section{Observations}~\label{sec:obs}
\subsection{Photometry}
Optical photometry of SN\,2023ixf was obtained with the 0.8\,m Tsinghua University-NAOC telescope (hereafter TNT; \citealt{Huang_TNT}) at Xinglong Observatory in China, the 1.5\,m AZT-22 telescope (hereafter AZT; \citealt{AZT}) at the Maidanak Astronomical Observatory (MAO) in Uzbekistan, the Lijiang 2.4\,m telescope (hereafter LJT; \citealt{LJT}) of Yunnan Astronomical Observatories in China, and the Schmidt 67/91\,cm Telescope (hereafter 67/91-ST) and 1.82\,m Copernico Telescope (hereafter Copernico) at the Asiago Astrophysical Observatory in Italy.
NIR photometry in the $JHK$ bandpasses was obtained with the Near Infrared Camera Spectrometer (NICS; \citealp{2001A&A...378..722B}) mounted on the 3.58\,m Telescopio Nazionale Galileo (TNG; \citealp{1994SPIE.2199...10B}) on the island of La Palma.
The photometry is provided in Table ~\ref{table:photometry}. All phases are given relative to the time of the first light estimated by~\citet{2024Natur.627..754L} (MJD 60082.788) throughout the paper.

\begin{table}[h]
\centering
\caption{Observed photometry of SN\,2023ixf}\label{table:photometry}%
\resizebox{\columnwidth}{!}{
\begin{tabular}{cccccc}
\hline
MJD & Phase (d) & Filter & Mag & err & Instrument \\
\hline
60086.612 & 3.824 & U & 10.064 &  0.064 & XLT \\
60086.612 & 3.824 & B & 11.128 &  0.051 & XLT \\ 
60086.612 & 3.824 & V & 11.376 &  0.048 & XLT  \\
60086.624 & 3.836 & g & 10.812 &  0.083 & XLT  \\
60086.624 & 3.836 & r & 11.413 &  0.046 & XLT  \\
60086.624 & 3.836 & i & 11.611 & 0.029 & XLT  \\
60087.000 & 4.212 & u & 10.926 &  0.113 & 67/91\,ST \\
60087.000 & 4.212 & g & 10.963 &  0.038 & 67/91\,ST \\
... & ... & ... &... & ... & ... \\
\hline
\end{tabular}}
\begin{flushleft} 
$Note:$ This table is available in its entirety in machine-readable form.
\end{flushleft}
\end{table}

Optical images obtained by all facilities were reduced following standard procedures, including bias subtraction and flat-field correction.
Images obtained by the TNT were processed using a custom ``ZURTYPHOT'' pipeline (Mo et al., in prep.). 
Reduction of the AZT, LJT, Copernico, and 67/91-ST images was carried out using the \textsc{AUTOPHOT} pipeline~\citep{AUTOPHOT}.
The World Coordinate System (WCS) was solved using \textsc{ASTROMETRY.NET} ~\citep{2010AJ....139.1782L}. 
We performed point-spread-function (PSF) photometry on the images; however, for accurate results, template subtraction might be required for photometry obtained at day 625 after the explosion. 

For all $BVgri$-band images, photometry is also performed to extract the instrumental magnitudes of any bright field stars that have photometric data available from the AAVSO Photometric All Sky Survey (APASS) DR9 Catalogue~\citep{2016yCat.2336....0H}.
By employing magnitudes of these local bright and isolated comparison stars, instrumental $BV$- and $gri$-band magnitudes of SN\,2023ixf were calibrated to the standard Johnson $BV$ system~\citep{1966CoLPL...4...99J} in Vega magnitudes and the Sloan Digital Sky Survey (SDSS) photometric system~\citep{1996AJ....111.1748F} in AB magnitudes, respectively. The instrumental $uz$-band magnitudes were calibrated using SDSS Release 16 \citep{SDSS16} to the standard SDSS photometric system. Instrumental $U$-band magnitudes of SN\,2023ixf were converted to  Johnson $U$ using the standard stars of~\cite{Zhang_2011fe}.
After the calibration of the instrumental magnitudes, we found that the $V$-band magnitudes from AZT show a systematic difference from those of other facilities by $\sim$0.15\,mag, likely caused by the deviation of the throughout from the standard system. Therefore, color-corrections have been applied to the $V$, $B$, and $U$ photometry~\citep{Landolt_1992_color_correction} from the AZT images. 

The NIR images were processed including flat-field and bias correction. Standard \texttt{IRAF}\footnote{{IRAF} is distributed by the National Optical Astronomy Observatories, which are operated by the Association of Universities for Research in Astronomy, Inc., under cooperative agreement with the U.S. National Science Foundation.}~\citep{1986SPIE..627..733T, 1993ASPC...52..173T} tasks were used to reduce the TNG/NICS images. Instrumental magnitudes were measured using SNOoPY\footnote{SNOoPy is a package for SN photometry using PSF fitting and/or template subtraction developped by E. Cappellaro. A package description can be found at \url{ https://sngroup.oapd.inaf.it/snoopy.html}}, and calibrated using the Two Micron
All Sky Survey (2MASS\footnote{\url{ http://irsa.ipac.caltech.edu/Missions/2mass.html/}}, \cite{2MASS}) catalog.

\subsection{Spectroscopy}
The spectroscopic campaign for SN\,2023ixf spans days $+$3 to $+$324. Optical observations  
were carried out using the following facilities. \\
(i) The Asiago Faint Object Spectrograph and Camera (AFOSC) on the 1.82\,m Copernico telescope operated by INAF Astronomical Observatory of Padova, atop Mount Ekar (Asiago, Italy). \\ 
(ii) The 1.22\,m Galileo Telescope (hereafter GT) equipped with the B$\&$C spectrograph at Osservatorio Astronomico di Asiago, Asiago, Italy. \\
(iii) the Beijing Faint Object Spectrograph and Camera (BFOSC) mounted
on the Xinglong 2.16\, m telescope (hereafter XLT;\citealt{XLT_2016_Zhang}), China. \\
(iv) The 2.4\,m LJT equipped with YFOSC (Yunnan Faint Object Spectrograph and Camera. \citealt{LJT_YFOSC}).\\

Seven NIR spectra were taken by the NICS instrument on the TNG, spanning the phase interval from days 4 to 93. 
Logs of the optical and NIR spectroscopy are presented in Tables~\ref{table:Log of optical spectroscopy} and \ref{table:Log of NIR spectroscopy}, respectively.

All spectra except for those from XLT were obtained with the long slit aligned at the parallactic angle~\citep{1982PASP...94..715F}. Spectra obtained with Copernico/AFOSC were reduced using the dedicated pipeline \textsc{Foscgui}. 
\footnote{
\textsc{Foscgui} is a graphical user interface (GUI) dedicated to extract the photometry and spectroscopy obtained with FOSC-like instruments. It was developed by E.~Cappellaro. A package description can be found at https://sngroup.oapd.inaf.it/foscgui.html
} 
Spectroscopic data obtained by other facilities were reduced utilizing standard \texttt{IRAF} routines including bias subtraction, flat-field correction, and removal of cosmic rays. 
Wavelength calibration was carried out using comparison-lamp exposures.
Flux calibration was performed using the sensitivity functions derived from the spectra of photometric standard stars observed during the same night, at airmasses similar to that of SN\,2023ixf.
Atmospheric extinction was corrected using the extinction curves of the observatories, and  telluric lines were removed using the standard-star spectra.

\section{Photometric Evolution} ~\label{sec:lc}
We estimate the Galactic reddening component along the SN\,2023ixf 
 line of sight as
$E(B-V)^{\rm MW}_{\rm 23ixf} = 0.008$\,mag by querying the NASA/IPAC NED\footnote{\url{https://ned.ipac.caltech.edu/}} Galactic Extinction Calculator, based on the extinction map derived from ~\citet{2011ApJ...737..103S}. Reddening due to the host galaxy was estimated as $E(B-V)^{\rm host}_{\rm 23ixf} = 0.032$\,mag according to the equivalent width of the Na\,{\sc\,I}\,D absorption lines measured from the high-resolution spectra of SN\,2023ixf~\citep{Smith_23ixf, Teja_23ixf, Zhang_23ixf} and the empirical relation derived between the line width and the dust reddening~\citep{2012MNRAS.426.1465P}. Thus, the total reddening is estimated as $E(B-V)^{\rm total}_{\rm 23ixf} = 0.04$\,mag. 

\begin{figure*}
    \centering
    \includegraphics[width=\textwidth]{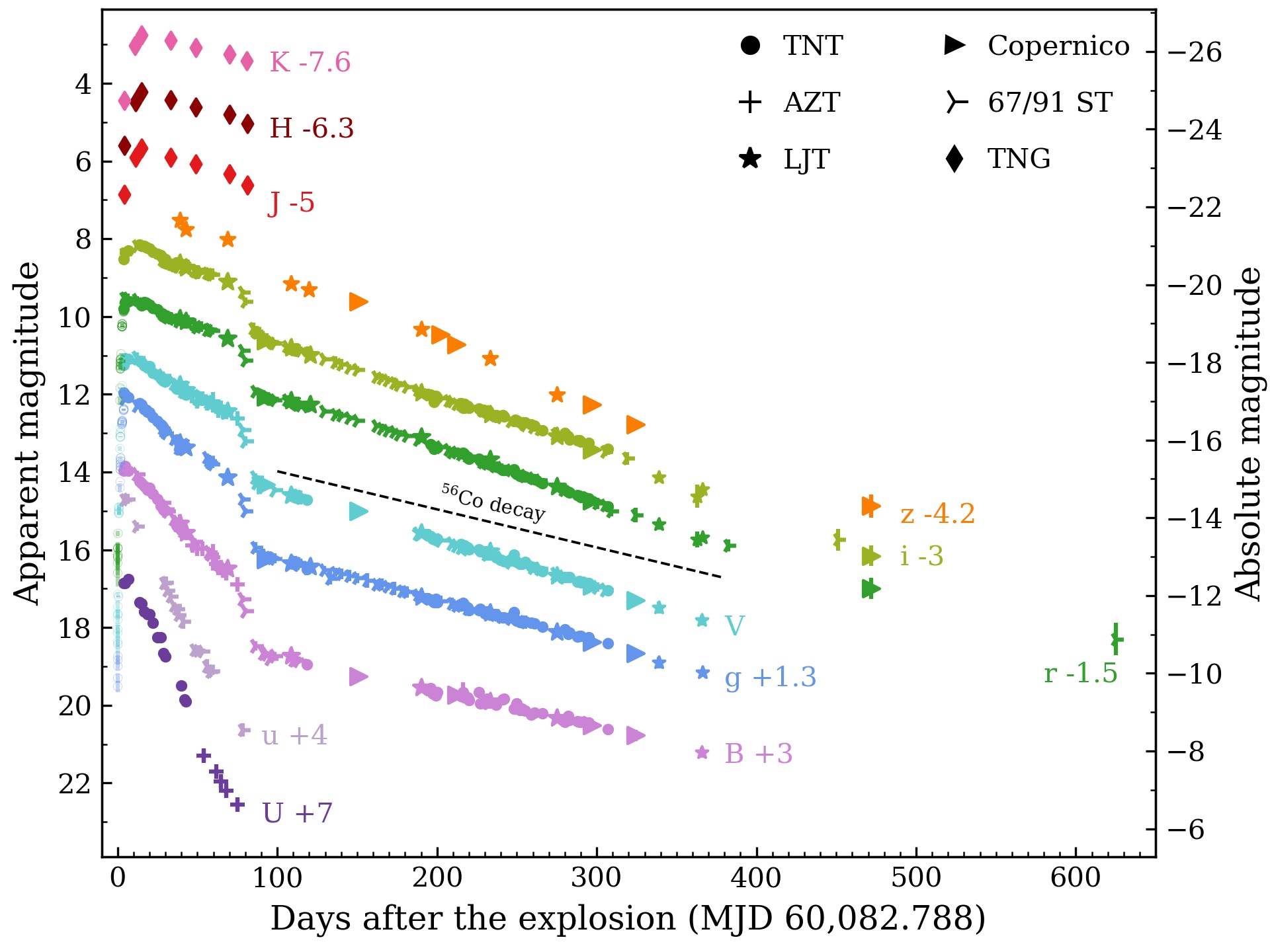}
    \caption{Optical and NIR light curves of SN\,2023ixf. Open circles overplot the early $g$-, $V$-, and $r$-band light curves from ~\cite{2024Natur.627..754L}. The black dashed line represents the decline expected for $^{56}$Co decay. 
    }
    \label{fig:photometry}
\end{figure*}

After correcting for the Galactic and host-galaxy extinction, the $UuBgVrizJHK$-band light curves of SN\,2023ixf are presented in Fig.~\ref{fig:photometry}, covering phases from 3 to 625 days after the explosion. A more thorough investigation of the photometric properties within the first hours to a few days is presented by~\cite{2024Natur.627..754L}.
The $V$ light curve of SN\,2023ixf indicates a rise of $\sim 6$ days before a peak magnitude of $M_V^{\rm peak}=-18.0\pm 0.1$ is reached. The high peak luminosity 
places SN\,2023ixf near the bright end of SNe II as illustrated by detailed sample analyses (e.g.,~\citealp{2019MNRAS.490.2799D}).

Following \citet{Anderson2014} and using the $V$ light curve of SN\,2023ixf, we derive the $s_1$ and $s_2$ parameters as $3.35 \pm 0.13$ and $1.93 \pm 0.10$\,mag per 100 days, respectively. The former and the latter describe the magnitude decline rate measured from the time between the peak luminosity and the start of the linearly declining plateau, and the magnitude decline rate when the SN has settled on its plateau phase, respectively. These derived results are consistent with those measured by~\citep{Zheng_23ixf}.
These values indicate an overall steeper decline compared to the corresponding mean values measured from the sample of~\cite{Anderson2014}, which are 2.65\,mag and 1.47\,mag per 100 days, respectively. 

After $\sim$70 days, the $V$ light curve of SN\,2023ixf shows a transition from the plateau phase to the radioactive tail. During the phase between 100 and 300 days after the explosion, the $V$-band decline rate is measured to be $1.23 \pm 0.01$\,mag\,(100\,d)$^{-1}$. Such a  rate appears to be faster compared to that expected for the $^{56}$Co$\rightarrow ^{56}$Fe decay, $\sim$0.98\,mag\,(100\,d)$^{-1}$, as shown in Fig~\ref{fig:photometry}. This indicates that $\gamma$-ray photons are not fully trapped inside the ejecta at this late phase. 

\begin{figure}
    \centering
    \includegraphics[width=1\columnwidth]{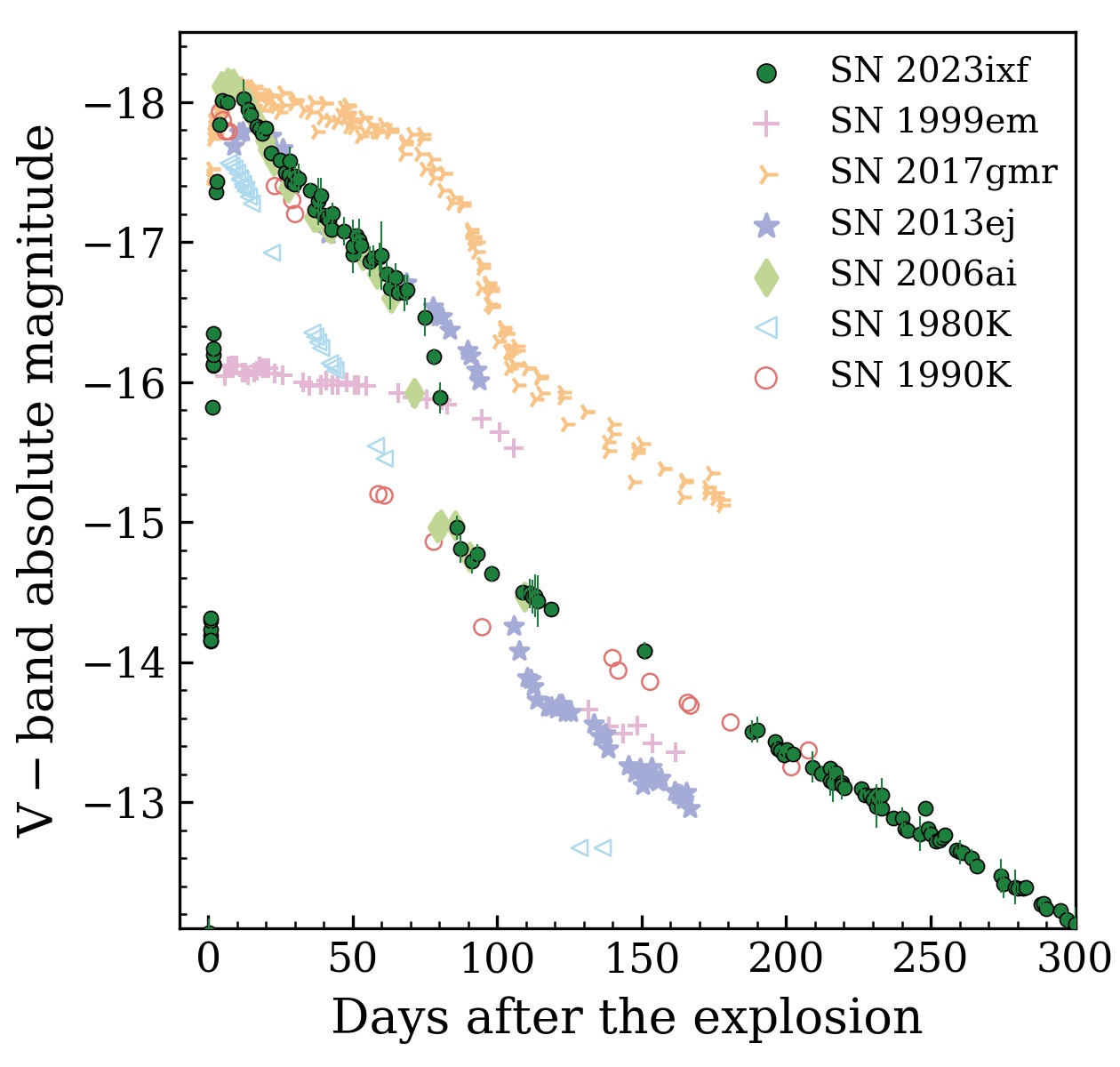}
    \caption{The absolute $V$-band light curve of SN\,2023ixf compared to those of the Type IIP/IIL SNe\,1999em \citep{2002PASP..114...35L}, 2017gmr~\citep{Jennifer_17gmr}, 
   2013ej \citep{2013ejHuang}, 2006ai \citep{Anderson2014}, 1980K~\citep{Buta_1980K}, and 1990K \citep{Cappellaro_1990K}. 
    All magnitudes have been corrected for the host and the Galactic extinction components.
    }
    \label{fig:photo_com}
\end{figure}

In Fig.~\ref{fig:photo_com}, we compare the absolute $V$ light curve of SN\,2023ixf with those of a selected sample of well-studied Type II SNe, including Type IIP SNe 1999em~\citep{2002PASP..114...35L} and 2017gmr~\citep{Jennifer_17gmr}, Type IIP/L SN 2013ej~\citep{2013ejHuang}, Type II short-plateau SN 2006ai~\citep{2021ApJ...913...55H}, and Type IIL SNe 1980K~\citep{Buta_1980K} and 1990K~\citep{Cappellaro_1990K}.
All light curves have been corrected for extinction in both the host galaxy and the Milky Way. 

The $V$ light curve of SN\,2023ixf shows a brighter absolute peak magnitude and shorter plateau compared to that of the prototypical SN IIP SN\,1999em. 
A shorter and less prominent plateau can also be identified in the light curves of SN 2006ai and Type IIP/IIL SN 2013ej~\citep{2015ApJ...806..160B}. The short-plateau Type II SN\, 2006ai, as discussed by~\citet{2021ApJ...913...55H}, is indicative of having a less massive H-rich envelope in its progenitor compared to that of normal SNe IIP \citep{Hillier_2019}. The $V$ light curves of Type IIL SNe\,1980K and 1990K exhibit a much faster and more linear post-peak decline than that of SN\,2023ixf during the plateau phase. Based on the comparison results discussed above, we suggest that SN\,2023ixf may be best described as a transitional object between Type IIP and IIL in terms of post-peak photometric evolution.

\section{Spectroscopic Evolution}~\label{sec:spec}
\subsection{Optical Spectroscopy}~\label{sec:spec_optical}
In Fig.~\ref{fig:spectra}, we present the spectral time series of SN\,2023ixf observed by XLT, Copernico, GT, NOT, and LJT from days 3 to 324.
All spectra are corrected for the redshift $z=0.000804$ of its host galaxy~\citep{Riess_distance} and dereddened adopting a total line-of-sight extinction $E(B-V)^{\rm total}_{\rm 23ixf} = 0.04$\,mag.
 
Narrow emission lines including H, He{\sc\,II}, and C{\sc\,IV} remain prominent within the first three days after the explosion. The ``flash features'' at high excitation states arise from the shock-ionized species in the CSM shell in close vicinity of the exploding progenitor.
This material was most likely expelled by pre-explosion instabilities in the years leading to the terminal SN explosion, and the spectral signatures reflect the surface composition of the dying progenitor and trace the explosion and shock physics of the SN.
Comprehensive analyses of the intranight spectroscopic evolution of SN\,2023ixf starting from day 1 has been carried out by \citet{Bostroem_23ixf}, \citet{Zheng_23ixf}, \citet{Daichi_23ixf}, \citet{Smith_23ixf}, \citet{Teja_23ixf}, and \citet{Zhang_23ixf}. 
As illustrated in Fig.~\ref{fig:spectra}, the flash features superimposed on the blue continuum persist for the first eight days.

Broad P~Cygni profiles emerged for spectral lines of Ba~{\sc ii}, Fe~{\sc ii}, Sc~{\sc ii} and Na~{\sc i} after the flash features were gone. A prominent H$\alpha$ P~Cygni line is seen after approximately day 20, indicating that the emission is dominated by the expanding H-rich envelope of the ejecta. Absorption on the blue side of H$\alpha$ at $\sim 13,000$\,km\,s$^{-1}$ appears at $\sim 20$\,d and disappears at $\sim 90$\,d. This notch was previously attributed to the high-velocity component of H$\alpha$~\citep{2002PASP..114...35L, Baron_2000_1999em} or Si~{\sc ii} $\lambda$6355~\citep{Pastorello2005_1, Valenti_2013ej}. 
The blue continuum diminishes as the SN enters the nebular phase. 

At about 90 days after explosion, the photosphere recedes through the H envelope. The absorption component of the P~Cygni profile of H$\alpha$ disappears and forbidden metal emission lines of [O~{\sc i}]\,$\lambda$5577, [O~{\sc i}] $\lambda\lambda$6300, 6364, [Ca~{\sc ii}] $\lambda\lambda$7291, 7323, and the Ca~{\sc ii} NIR triplet emerge and start to dominate the spectra. 
Na~{\sc i} $\lambda\lambda$5890, 5896, [O~{\sc i}] $\lambda\lambda$6300, 6364, H$\alpha$, and [Ca~{\sc ii}] $\lambda\lambda$7291, 7323 all show multipeak profiles after about day 100, as discussed in detail in Section~\ref{sec:lines profile}.

\subsection{Comparison with Other SNe II}
\begin{figure*}[h]
    \centering
    \includegraphics[width=\textwidth]{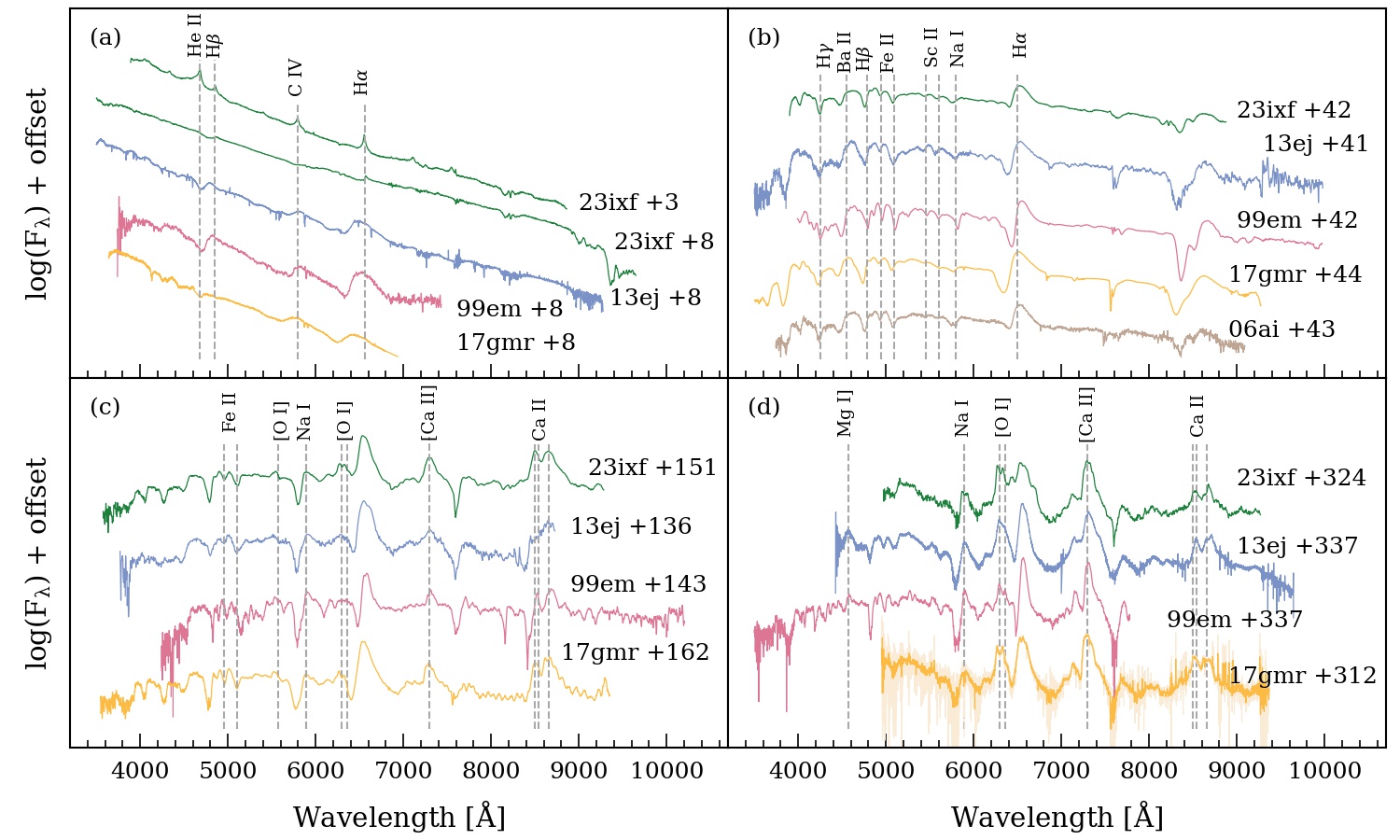}
    \caption{Spectra of SN\,2023ixf at days 5 (a), 40 (b), 151 (c), and 330 (d) compared to that of the Type IIP/L SNe\,2013ej~\citep{Valenti_2013ej,Yuan_2013ej, 2013ejHuang}, 1999em~\citep{2002PASP..114...35L}, 2017gmr~\citep{Jennifer_17gmr} (the spectrum at day 312 is binned, and the original version is plotted with a fainter color), and 2006ai~\citep{2021ApJ...913...55H} at similar phases. In each panel, major spectral features are marked by vertical dashed lines.
    For better display, all spectra were shifted arbitrarily and presented with a logarithmic scale.}
    \label{fig:comparison}
\end{figure*}

Fig.~\ref{fig:comparison} compares
the spectra of SN\,2023ixf with those of other well-studied SNe IIP/L at similar epochs, namely SNe\,2013ej\footnote{This previously unpublished spectrum at 337\,d is from A. V. Filippenko's group at UC Berkeley. It was obtained on 26 June 2014 with the DEep Imaging Multi-Object Spectrograph(DEIMOS; \citealt{Faber2003}) on the 10\,m Keck II telescope on Maunakea, with the long slit aligned at the parallactic angle \citep{1982PASP...94..715F}. The spectrum was reduced with standard \texttt{IRAF} routines~\citep{Silverman2012} and was flux calibrated using spectrophotometric standard stars.}~\citep{Valenti_2013ej, Yuan_2013ej, 2013ejHuang}, 1999em~\citep{2002PASP..114...35L}, 2017gmr~\citep{Jennifer_17gmr}, and 2006ai~\citep{2021ApJ...913...55H}. All spectra were corrected for the redshift, as well as for host-galaxy and Galactic extinctions. 

As presented in Fig.~\ref{fig:comparison}(a), while other comparison SNe have already developed broad P~Cygni profiles of the Balmer lines, the spectrum of SN\,2023ixf is still characterized by weak flash lines atop a blue continuum. The long-lived narrow flash features 
indicate that the process of the ejecta engulfing the CSM persists for the first eight days. Such an extended flash-ionization phase is significantly longer compared to the $\sim 5$\,day maximum duration observed in most  CCSNe~\citep{2023ApJ...952..119B}, 
indicating a more radially extended CSM shell around the progenitor and more extensive mass loss before the explosion compared to other cases.

Fig.~\ref{fig:comparison}(b) shows a comparison of the spectrum of SN\,2023ixf at day 42 with the other SNe at similar phases.
SN\,2023ixf presents a shallower H$\alpha$ P~Cygni absorption than others. \cite{Gutierrez2014, Gutierrez2017} found that the smaller ratio of absorption to emission of H$\alpha$ is correlated with brighter and faster declining light curves. 
The shallower absorption component may imply a deficit in absorbing material at high velocities, which may naturally be reproduced by a rather steep density gradient in the H envelope. A lower envelope mass retained before the explosion may also contribute to the smaller ratio of the absorption to emission of H$\alpha$~\citep{Gutierrez2014, Gutierrez2017, 2014MNRAS.445..554F, 2021ApJ...913...55H}. 
\cite{Hillier_2019} proposed that the interaction between the ejecta and CSM could also produce a weak or no H$\alpha$ absorption. In addition, SN\,2023ixf has fewer and shallower metal lines than SN\,1999em.

Comparison of the day 150 spectrum of SN\,2023ixf with those of other SNe is presented in Fig.~\ref{fig:comparison}(c).
While others still exhibit a P~Cygni profile of H$\alpha$, the H$\alpha$ absorption has already disappeared in SN\,2023ixf. 
SN\,2023ixf has more prominent [O~{\sc i}] $\lambda$5577, [O~{\sc i}] $\lambda\lambda$6300, 6364, and [Ca~{\sc ii}] $\lambda\lambda$7291, 7323 lines compared to other SNe. The early appearance and strong emission of [O~{\sc i}] may hint that its progenitor was partially stripped before the explosion \citep{Elmhamdi_2011_O}. 
The Ca~{\sc ii} NIR triplet of SN\,2023ixf is comparable to that of SN\,2017gmr but stronger than that of SN\,1999em.

The comparison of the day 328 spectrum is shown in Fig.~\ref{fig:comparison}(d). 
Both SN\,2023ixf and SN\,2017gmr exhibit a double-peaked [O~{\sc i}] $\lambda\lambda$6300, 6364 profile, but SN\,2013ej has single-peaked [O~{\sc i}]. 
While the line profiles of SN\,2023ixf, SN\,2013ej, and SN\,2017gmr are overall similar, their light curves show significant differences from each other (Fig.~\ref{fig:photo_com}).

\subsection{Metallicity}
\begin{figure}[h]
    \centering    \includegraphics[width=\columnwidth]{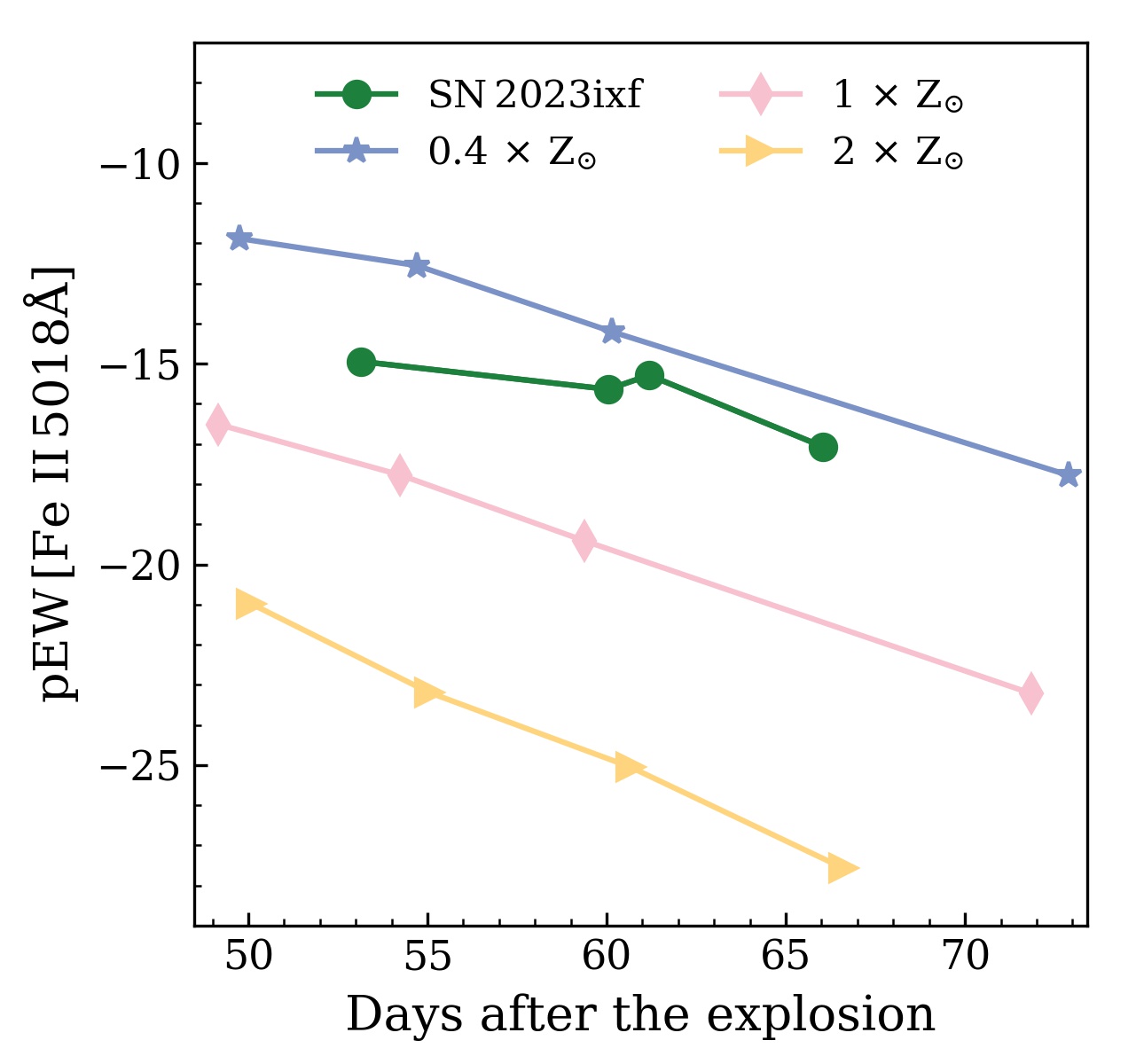}
    \caption{The pEW of Fe~{\sc ii} $\lambda$5018 in SN\,2023ixf compared with the models presented by~\citet{2014MNRAS.440.1856D}. The 0.4, 1, and 2 $\times$ Z$_{\odot}$ models are shown by the color-coded lines (see legend).}
    \label{fig:metallicity}
\end{figure}

The line strengths of some metals measured during the photospheric phase were found to be correlated with the metallicity of their progenitors for SNe IIP/IIL~\citep{2014MNRAS.440.1856D}. 
For instance, the pesudo-equivalent width (pEW) of Fe~{\sc ii}\,$\lambda$5018 can be regarded as a metallicity indicator. 
The measured pEWs of Fe~{\sc ii}\,$\lambda$5018 of SN\,2023ixf from the spectra between approximately days 50 and 70 and those of the models are presented in Fig.~\ref{fig:metallicity}. The time interval was chosen to represent the mid-to-late plateau phase, during which the photosphere still probes the H envelope of the progenitor~\citep{2014MNRAS.440.1856D}. Without contamination through the outward chemical mixing from the inner core, the strengths of metal lines measured at the outer part of the SN ejecta can reflect the chemical content of the progenitor.
As the photosphere of SN\,2023ixf recedes into its inner He-rich core after $t \approx 70$\,days as indicated by the termination of the plateau phase (Fig.~\ref{fig:photometry}), comparison of the observations with the model can be made up to about day 70.

From Fig.~\ref{fig:metallicity}, one can see that the metallicity of SN\,2023ixf falls into the range (0.4--1)\,Z$_{\odot}$. We remark that one should remain cautious about the inferred range of the metallicity, as the reference models were constructed for SNe IIP with a prominent light-curve plateau, while both the photometric and spectroscopic properties of SN\,2023ixf show close resemblances with Type IIL SNe.

\subsection{Expansion Velocity}~\label{sec:velocity}
\begin{figure}[h]
    \centering    
    \includegraphics[width=\columnwidth]{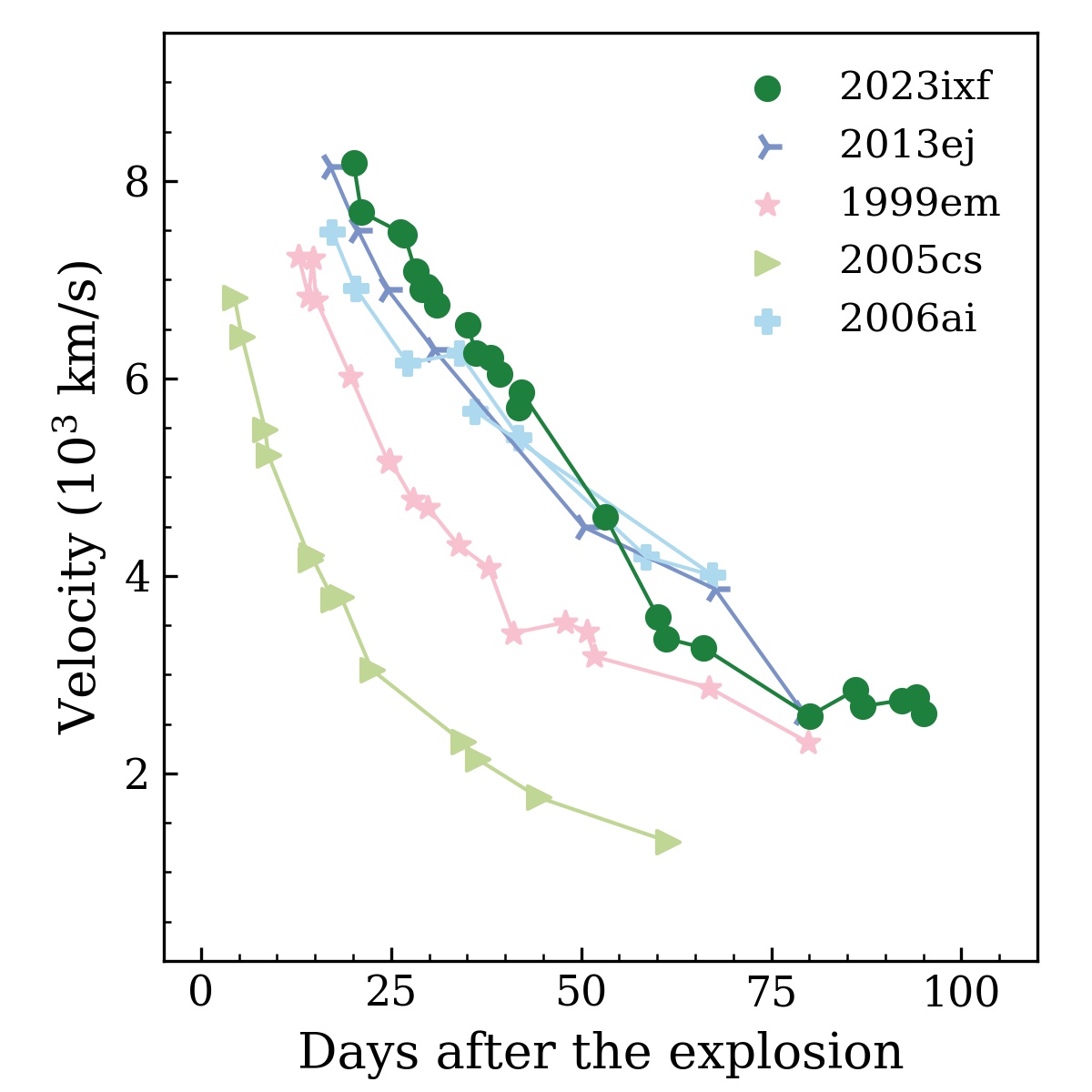}
    \caption{Comparison of the evolution of the Fe~{\sc ii} $\lambda$5169 velocity of SN\,2023ixf with those of SNe\,2013ej, 1999em, 2005cs, and 2006ai.}
    \label{fig:velocity_comparison}
\end{figure}

Simulations of SN atmospheres based on the non-LTE CMFGEN model~\citep{1998ApJ...496..407H} suggest that the photospheric velocity of SNe II can be well traced by the absorption minimum of  the Fe{\sc\,ii}\,$\lambda$5169 feature ($v_{\rm Fe}$; \citealp{2005A&A...439..671D}).
We measured the absorption minima of Fe~{\sc ii} $\lambda$5169 in SN\,2023ixf. At $t \approx 100$\,d, the profile of Fe~{\sc ii} $\lambda$5169  no longer follows a Gaussian profile, perhaps owing to line blending, so the velocity is not measured thereafter. 
The velocity evolution of SN\,2023ixf is shown in Fig.~\ref{fig:velocity_comparison}, together with those measured for the Type IIP SNe\,1999em~ \citep{2012MNRAS.419.2783T} and 2005cs~\citep{2012MNRAS.419.2783T}, the Type IIP/IIL SN\,2013ej~\citep{2013ejHuang}, and the Type II short-plateau SN\,2006ai~\citep{2021ApJ...913...55H}. 

Comparison of the Fe{\sc\,ii} $\lambda$5169 velocity evolution of SN\,2023ixf and other Type IIP/L SNe suggests that the former also follows an exponential-like decline as observed in the other presented cases. 
SN\,2023ixf displays an ejecta velocity evolution similar to that measured for SNe\,2013ej and 2006ai, significantly higher compared to that of the normal SN IIP\,1999em~\citep{Elmhamdi_1999em, Utrobin_2017} and the subluminous SN IIP \,2005cs~\citep{Pastorello2005_1, Pastorello2009}. 
The distinct velocities between subluminous and normal SNe II may be indicative of the associated total energy and the debris of the SN explosion. For example, the former and the latter groups are thought to be linked to the O-Ne-Mg core and Fe core from progenitors with lower ($\sim$8--10\,M$_{\odot}$) and higher masses ($\gtrsim$10\,M$_{\odot}$), respectively~\citep{Fraser2011, Janka2012}. Faster expansion velocities of SNe IIP are considered to be linked with higher explosion energies~\citep{Dessart_2010MNRAS.408..827D, Gutierrez2017}.

\subsection{Distance}
The expanding photosphere method (EPM) provides an independent estimate of the distance to a SN based on a comparison between the physical and angular radii of its photosphere, denoted by $r$ and $\theta$, respectively~\citep{1974ApJ...193...27K}. 
The former can be calculated by multiplying the expansion velocity of the photosphere ($v_{\rm phot}$) by the time elapsed since the SN explosion ($t-t_{0}$), and the latter can be used to estimate the distance to the SN ($D$) through direct geometric calculation, 
\begin{equation}
\frac{v_{\rm phot}(t - t_0)}{D} = \theta\, .
\label{equ:epm}
\end{equation}

We adopt the velocities measured from the absorption minimum of Fe\,{\sc ii}\,$\lambda5169$ to represent $v_{\rm phot}$ as discussed in Section~\ref{sec:velocity}.
Following the prescriptions described by~\citet{2002PASP..114...35L} and~\citet{ 2005A&A...439..671D}, the comparison between the synthetic magnitude calculated from a blackbody spectrum and the photometry of the SN in a given bandpass can be facilitated by minimizing the quantity $\epsilon$, and $\theta$ can be derived via 
\begin{equation}
\resizebox{0.9\columnwidth}{!}{$\epsilon = \mathop{\sum}\limits_{BVI} {\left\{m_{\rm BVI} - A_{\rm BVI} + 5\,\mathrm{log}\,\theta + 5\,\mathrm{log}\,[\zeta(T_{\rm c})] - b_{\rm BVI}(T_{\rm c})\right\}}^2$}\, ,
\end{equation}
where $b_{\rm BVI}$ is the synthetic broadband magnitude of the Planck function with temperature $T_{\rm c}$.
The dilution factor $\zeta$ corrects the difference between the thermalization and photospheric radii whereas the former is determined by the blackbody temperature~\citep{1996ApJ...466..911E, 2001ApJ...558..615H}.
We transform the $i$-band magnitude to Johnson $I$ with the Lupton (2005) color transformations\footnote{\url{https://www.sdss3.org/dr8/algorithms/sdssUBVRITransform.php\#Lupton2005.}}.

Minimization of the quantity $\epsilon$ was achieved through a Markov Chain Monte Carlo (MCMC) approach. Upon the determination of $\theta$ and $v_{\rm phot}$, the distance to the SN was then carried out based on Equation~\ref{equ:epm}.
The priors of parameters are assumed to be uniform.
\cite{2009ApJ...696.1176J} show a clear departure from linearity between $\theta/v$ versus $t$ after $t \approx 45$--50 days, likely caused by the progressively emerging spectral lines that deviate the continuum emission of a blackbody~\citep{2005A&A...439..671D}.
Therefore, we restricted the EPM analysis to before this phase. 

Based on the EPM approach, we estimate the distance to M\,101 as $D=6.35^{+0.31}_{-0.39}$\,Mpc ($\mu=29.01^{+0.10}_{-0.14}$\, mag). The distance derived using the standard candle method~\citep{Zheng_23ixf} is $28.67 \pm 0.14\,$mag.
Distances measured with Cepheids range from 6.19 to 8.99\,Mpc~\citep{6.19Mpc, 8.99Mpc}, while those measured with SNe Ia range from 5.92 to 7.52\,Mpc~\citep{5.92Mpc, 7.52Mpc}. Therefore, our result is consistent with distances measured from other methods.

\subsection{H$\alpha$ Profile During the Nebular Phase} ~\label{sec:Ha profile}

\begin{figure}[h]   
    \centering    \includegraphics[width=0.5\textwidth]{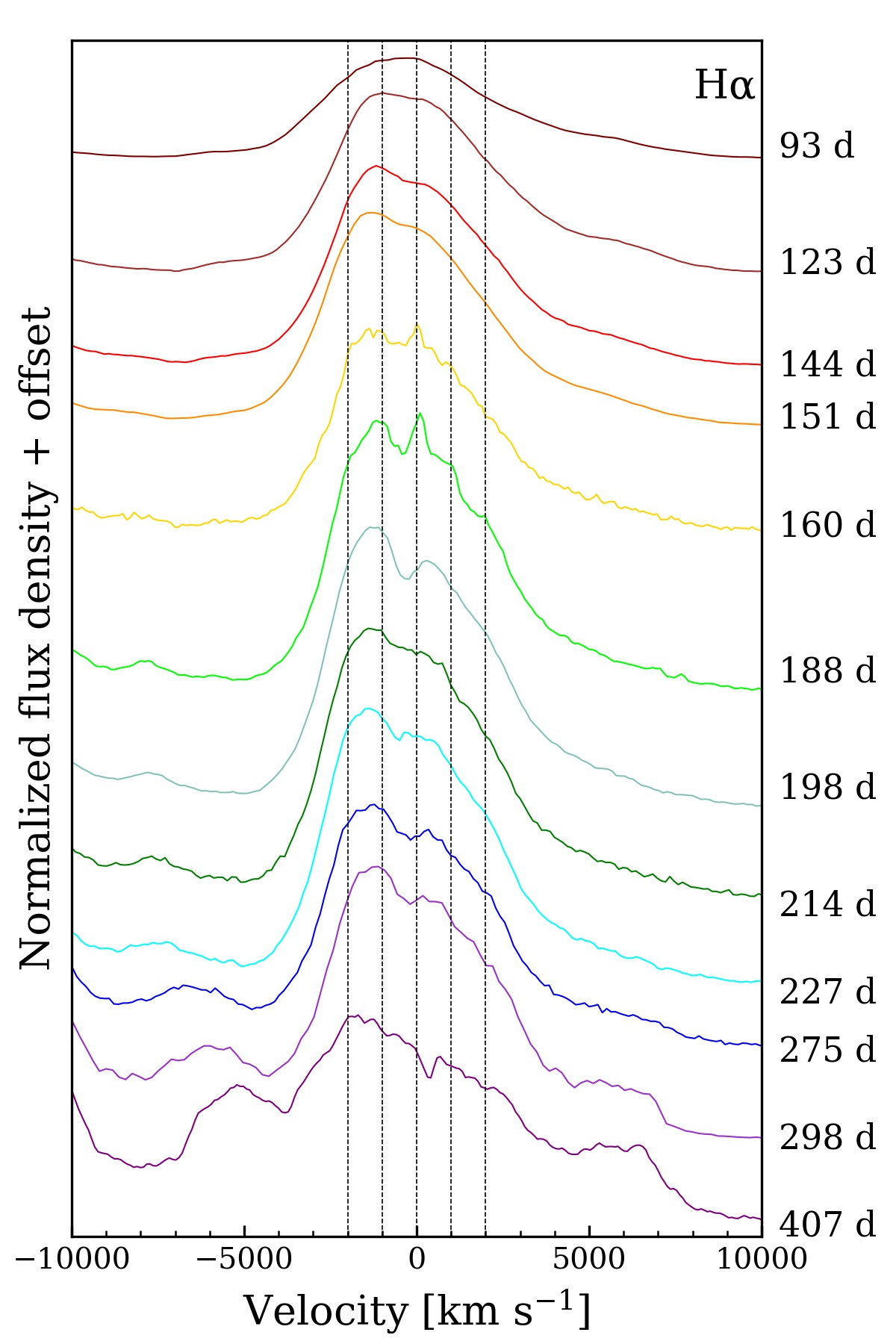}
    \caption{Temporal evolution of the H$\alpha$ profile from days 93 to 298. 
    Three vertical black dashed lines mark the rest-frame velocities of $-$2000, $-$1000, 0, $+$1000 and $+$2000\,km\,s$^{-1}$, respectively.}
    \label{fig:Ha}
\end{figure}

During the photospheric phase, the H$\alpha$ line of SN\,2023ixf is characterized by a P~Cygni profile, as shown in Fig.~\ref{fig:spectra}. After $t \approx 90$ days, the absorption component of H$\alpha$ is diminished and the emission component of H$\alpha$ starts to develop into a double-peaked profile. 
The evolution of the nebular-phase H$\alpha$ profile, normalized by the pseudocontinuum, is shown in Fig.~\ref{fig:Ha} (note that the 407\,day spectrum has been published by~\cite{Zheng_23ixf}). It is evident that the H$\alpha$ line of SN\,2023ixf developed a complex profile when entering the nebular phase. For instance, at $t \approx 189$\,days, the profile exhibits a multipeak structure with emission peaks ranging from $\sim -1500$\,km\,s$^{-1}$ (blueshifted) to $\sim +1500$\,km\,s$^{-1}$ (redshifted). 
The blueshifted component appears stronger than the redshifted counterpart. Such an asymmetric H$\alpha$ profile was also observed in other SNe II, such as SN 1999em~\citep{2002PASP..114...35L}, SN\,2013ej~\citep{2013ejHuang}, and SN\,2017gmr~\citep{Jennifer_17gmr}, and has been attributed to bipolar distribution of Ni.

The suppressed redshifted component could instead be attributed to dust formation,
as first proposed for the late-time spectral evolution of SN\,1987A~\citep{Lucy_1987A}. On-site dust formation has also been reported in some CCSNe, such as SN\,2006jc \citep{Smith_2006jc}, SN\,1999em \citep{Elmhamdi_1999em}, SN 2010jl~\citep{Zhang_2012}, and SN 2018hfm~\citep{Zhang_2022}, most likely taking place in a cold dense shell (CDS) between the forward-shock and reverse-shock fronts~\citep{2001MNRAS.326.1448C, 2004MNRAS.352.1213C, 2009MNRAS.394...21D}.
In SN 2023ixf, further evidence of dust formation can be inferred from the flux excesses in the NIR and mid-infrared (MIR) light curves taken beyond $\sim$120 days~\citep{Singh_2023ixf, Van_dust}. In particular, the MIR light curve displays a prominent brightening at day $\sim$210 at $\sim$4.6\,$\mu$m, which can be explained by the emergence of line emission from the 1--0 vibrational band of carbon monoxide (CO) at 4.65\,$\mu$m.

We also note that starting from $t \approx 90$\,days, a notch on the red shoulder of the H$\alpha$ line started to develop, indicating the contribution from a flat-topped component underlying the emission feature (Fig.~\ref{fig:spectra}). 
As shown by the $t \approx 93$\,day spectrum in Fig.~\ref{fig:Ha}, the box-shaped continuum extends to a velocity range of $\sim 5000$--8000\,km\,s$^{-1}$ with respect to the rest-frame zero velocity of H$\alpha$. 
Such a boxy-shaped emission feature provides signatures of strong interaction between the expanding ejecta and shell-like CSM. The edges of the emission indicate the expanding velocity of the ejecta (see, e.g., \citealp{2022A&A...660L...9D}). Thus, this boxy-shaped emission suggests a CSM shell at a distance of $\sim6.2 \times 10^{15}$\,cm. Assuming the CSM shell was produced by a typical RSG wind at a velocity of $\sim$10\,km\,s$^{-1}$, this material was expelled $\sim$\,200\,yr before the explosion. A blueshifted peak at $\sim -8000$\,km\,s$^{-1}$ started to emerge at day 190, which shifted gradually to $\sim -5000$\,km\,s$^{-1}$ by $t \approx 400$ days.
The different profiles on the right and left sides of H$\alpha$ above 5000\,km\,s$^{-1}$ may naturally be attributed to the presence of a spherically asymmetric dense shell. 
Furthermore, we note that the shoulder on the right side of the H$\alpha$ profile persisted until day $\sim$150, and emerged again from roughly day 275.  
This may indicate a persisting interaction with multiple CSM shells. The latter may result from multiple episodes of mass loss that lead to  SN\,2023ixf at $\gtrsim$ 200\,yr before the expolsion. These bumps might also be attributed to bipolar ejecta at the largest velocity interacting with the CSM~\citep{Smith2010jp}.

The narrow component at 0\,km\,s$^{-1}$ appearing at days 160--189 might be from the host galaxy, as weak [S~{\sc II}] $\lambda\lambda$6716, 6731 lines are observed at the same epoch. 

\subsection{Temporal Evolution of the Spectral Line Profiles in the Nebular Phase}
~\label{sec:lines profile}
\begin{figure}   
    \centering    \includegraphics[width=0.5\textwidth]{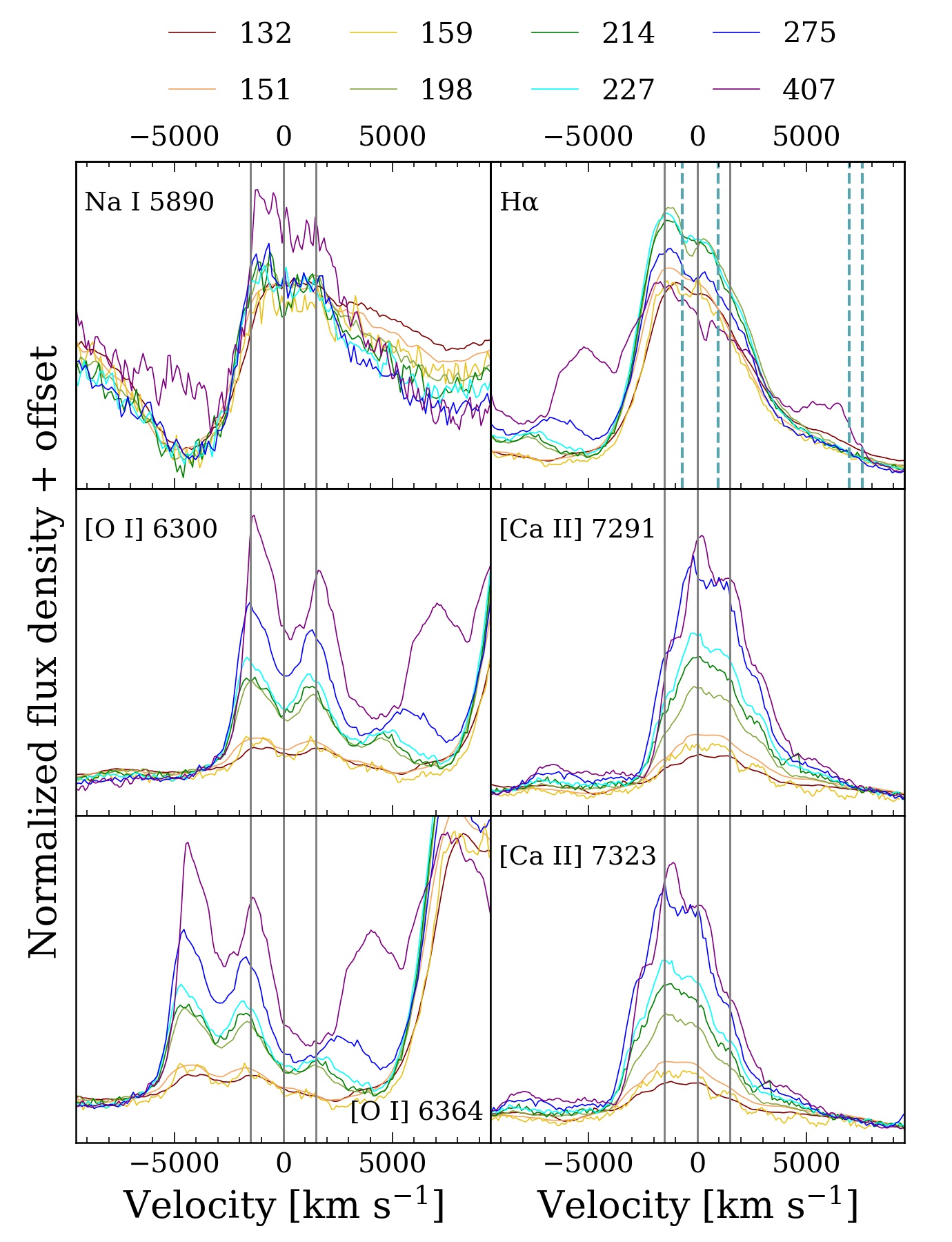}
    \caption{
    Temporal evolution of the spectral line profiles centered at the rest-frame wavelengths of Na\,{\sc i}\,$\lambda$5890, 5896, [O\,{\sc i}]\,$\lambda$6300, [O\,{\sc i}]\,$\lambda$6364, H$\alpha$, [Ca\,{\sc ii}]$\lambda$\,7291, and [Ca\,{\sc ii}]$\lambda$\,7323 as labeled in each subpanel. All spectra are presented in velocity space. Three vertical black dashed lines mark the rest-frame velocities of $-$1500, 0, and $+$1500\,km\,s$^{-1}$, respectively.
    The blue dotted lines mark the [N~{\sc ii}] $\lambda\lambda$6548, 6583 and [S~{\sc ii}] $\lambda\lambda$6716, 6731 emission lines from an H~{\sc ii} region in the host galaxy.
    }
    \label{fig:all_ele}
\end{figure}

For the purpose of investigating the temporal evolution of the morphology of several prominent spectral features in the nebular phase of SN\,2023ixf, we 
normalized the spectra by the pseudocontinuum. Spectral lines of Na~{\sc i} $\lambda\lambda$5890, 5896, [O~{\sc i}] $\lambda\lambda$6300, 6364, H$\alpha$, and [Ca~{\sc ii}] $\lambda\lambda$7291, 7323, spanning days 132 to 407, are shown in Fig.~\ref{fig:all_ele}.
We inspect the line velocities of [O{\sc}~{\sc i}] $\lambda\lambda$6300, 6364 and [Ca{\sc}~{\sc ii}] $\lambda\lambda$7291, 7323 with respect to their rest-frame wavelengths.

First, as the photosphere progressively recedes, the emission component of the Na\,{\sc i}\,D $\lambda\lambda$5890, 5896 doublet emerges in the spectra.
A double-peaked emission profile centered at  zero velocity can be identified, with a nearly constant pEW over time. The blueshifted and redshifted components peak at $\sim $-1500 and $+$1500\,km\,s$^{-1}$, respectively. As lines from [O~{\sc i}] and [Ca~{\sc ii}] strengthen with time with respect to the pseudocontinuum, H$\alpha$ diminishes gradually.
Second, the H$\alpha$ line also exhibits a dual-peaked morphology as seen from the Na~{\sc i} $\lambda\lambda$5890, 5896 doublet, which also peaked at $\sim-$1500 and $+$1500\,km\,s$^{-1}$, respectively.
Third, the velocity profiles centered at 6300 and 6364\,\AA\ suggest that both doublets display a blueshifted component at $-$1500\,km\,s$^{-1}$. A similar characterization can also be inferred for the [Ca~{\sc ii}] $\lambda\lambda$7291, 7323 profile. 
Finally, a redshifted component can be identified from the shoulders in the red-side profiles of the [Ca~{\sc ii}] doublet at a velocity of $\sim+$1500\,km\,s$^{-1}$. We tentatively identify a redshifted component in the [O~{\sc I}] lines as they are likely blended with the blue wing of H$\alpha$.

In summary, the zoom-in of the pseudocontinuum-normalized profiles of the presented lines of interest exhibits 
a more prominent component shifted by $\sim$1500\,km\,s$^{-1}$ to the blue, and another relatively weaker component shifted by $\sim$1500\,km\,s$^{-1}$ to the red. The dual-peaked spectral line profiles of SN\,2023ixf might hint at an aspherical distribution of the ejecta~\citep{Chugai_17gmr, 2005AstL...31..792C}.

\subsection{Near-Infrared Spectroscopy}
\begin{figure*}
    \centering
    \includegraphics[width=\textwidth]{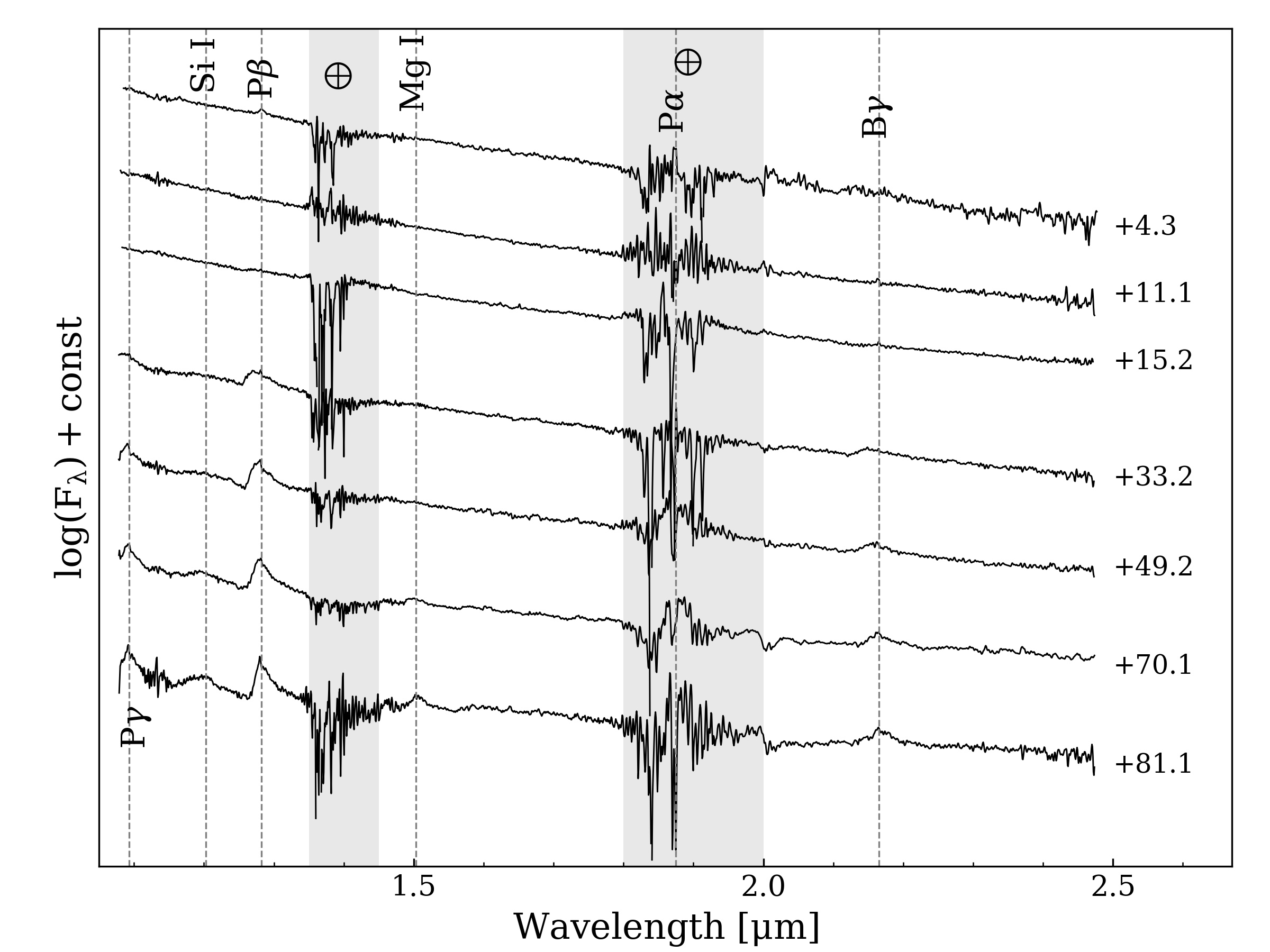}
    \caption{NIR spectra of SN\,2023ixf observed by TNG from $t \approx 4$ to 80 days after the explosion. Phases are indicated to the right of each spectrum. Several prominent spectral lines are marked with vertical dashed lines. Telluric lines are shown by crossed circles and gray shade.}
    \label{fig:NIR_spectra}
\end{figure*}

In Figure~\ref{fig:NIR_spectra}, we present a total of seven NIR spectra of SN\,2023ixf obtained with TNG+NICS.
At $\sim$4 days past explosion, the spectrum exhibits a weak narrow emission line of P$\beta$\,1.282\,$\mu$m superimposed on the continuum. Seven days later, P$\beta$ disappears and no prominent line can be identified. At $\sim 30$ days after explosion, P$\beta$ and 
B$\gamma$\,2.165\,$\mu$m appear and strengthen over time. P$\beta$ develops a P~Cygni profile at 49\,d.
Mg~{\sc i}\,1.503\,$\mu$m and Si~{\sc i}\,1.203\,$\mu$m appear at about 70\,d.
P$\alpha$\,1.875\,$\mu$m is seriously compromised by the telluric absorption. No CO overtone at 2.3\,$\mu$m is discernible.

\begin{figure}[h]
    \centering
    \includegraphics[width=\columnwidth]{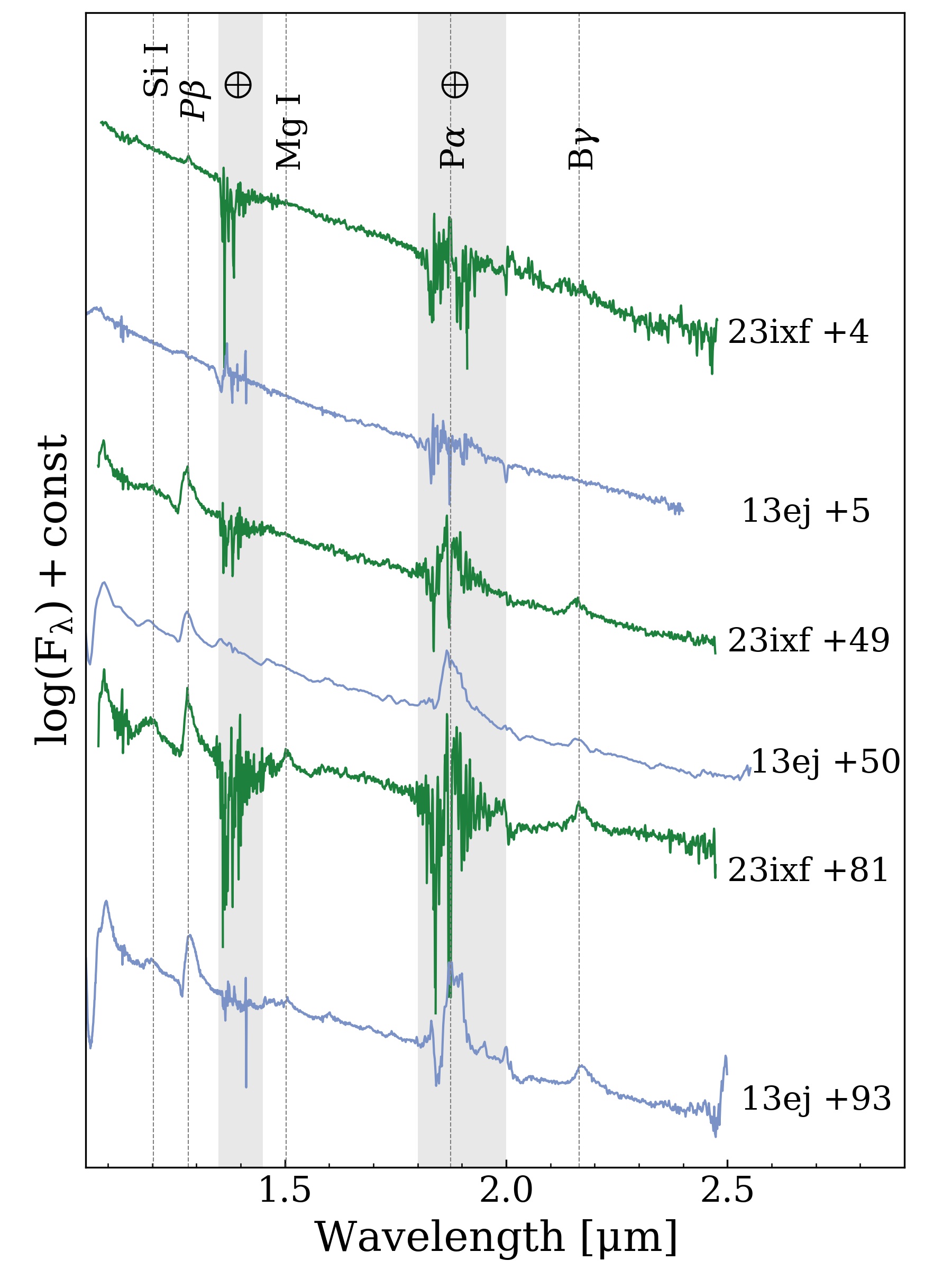}
    \caption{Comparison of the NIR spectra of SN\,2023ixf with those of SN\,2013ej at similar phases. Several prominent spectral features are marked by dashed lines, and telluric absorption is shown by crossed circles and gray shade.}
    \label{fig:NIR_comparion}
\end{figure}

We compare the NIR spectra of SN\,2023ixf with those of SN\,2013ej~\citep{2019ApJ...887....4D} at similar phases (Fig.~\ref{fig:NIR_comparion}). The narrow P$\beta$ observed in SN\,2023ixf at $t \approx 4$ days cannot be discerned in SN\,2013ej at $t \approx 5$ days. The spectra of SN\,2023ixf show much stronger and broader Mg~{\sc i}\,1.503\,$\mu$m and Si~{\sc i}\,1.203\,$\mu$m than SN\,2013ej.

\section{Discussion}~\label{sec:discussion}

\subsection{$^{56}\text{Ni}$ mass}\label{sec:Ni mass}
$^{56}\text{Ni}$ is synthesized during the SN explosion. The trapping of the $\gamma$-rays produced in the decay chain $^{56}\text{Ni}\rightarrow ^{56}\text{Co}\rightarrow ^{56}\text{Fe}$ is the main power source of the radioactive-tail luminosity of SNe. Therefore, the bolometric flux of the tail can be utilized to estimate the amount of radioactive $^{56}$Ni synthesized in the SN explosion.

During the plateau and post-plateau phases, the $V$-band light curve of Type II/IIP SNe provides a good proxy for the bolometric evolution. This has been characterized in detail by \citet{2009ApJ...701..200B}, who found a root-mean-square (rms) scatter of $\sim$0.11\,mag in the difference between the $V$ and bolometric light curves:
\begin{equation}
\resizebox{0.9\columnwidth}{!}{$\mathrm{log} L_{\rm t} =\,-0.4[BC+ V_{\rm t}\,-\,A(V)\,-\,11.64]\,+\,\mathrm{log}_{10}(4\,\pi\,D^2)\, ,$}
\label{equ:V to bolometric}
\end{equation}
where $L_{\rm t}$ is the SN luminosity in units of erg\,s$^{-1}$ when it settles to the radioactive-decay tail, $D$ is the distance in cm, and BC denotes the bolometric correction which has been estimated as $-0.70 \pm 0.02$~\citep{2009ApJ...701..200B}.

As shown in Fig.~\ref{fig:photometry} and discussed in Section~\ref{sec:lc}, the radioactive tail of SN\,2023ixf declines faster than the decay of $^{56}\text{Co}$ assuming full trapping of gamma-ray photons.
Therefore, we infer an incomplete trapping of $\gamma$-ray photons, which could be attributed to the decreasing photon diffusion time as the SN ejecta expand and become more transparent.
Accounting for the $\gamma$-ray leakage, the luminosity from solely the $^{56}$Ni$\rightarrow ^{56}$Co$\rightarrow ^{56}$Fe decay chain can be fitted following the prescription of \citet{Yuan_2013ej}:
\begin{equation}
\resizebox{0.9\columnwidth}{!}{$L\,=\,1.41\,\times\,10^{43}\,m_{\rm Ni}\,(e^{-t/t_{\rm Co}}\,-\,e^{-t/t_{\rm Ni}})(1-\,e^{{t_1}^2/t^{2}})\, ,$}
\label{equ:Ni decay}
\end{equation}
where $L$ is the bolometric luminosity, 
$m_{\rm Ni}$ gives the mass of $^{56}$Ni synthesized in the SN explosion,
$t$ represents the time since explosion, and $t_{\rm Co}$ and $t_{\rm Ni}$ are the $e$-folding times of $^{56}\text{Co}$ and $^{56}\text{Ni}$ (111.4 and 8.8\,days, respectively). The term $t_{1}$ represents the characteristic time when the $\gamma$-ray optical depth of the ejecta drops to unity.
This characteristic timescale is determined by the average $\gamma$-ray opacity, the total mass of the ejecta, and the kinetic energy of the SN. The $t^{-2}$ dependence of the optical depth is due to the decreasing column density as the SN ejecta undergo homologous expansion ~\citep{Clocchiatti_1997}.

We therefore adopt Equation~\ref{equ:V to bolometric} to convert the $V$-band photometry of SN\,2023ixf between days 90 and 350 to the bolometric light curve and estimate the mass of $^{56}$Ni synthesized in the ejecta by fitting Equation~\ref{equ:Ni decay} assuming incomplete trapping of X-rays and electrons/positrons.
The best fit to the single radioactive decay chain model suggests 
$m_{\rm Ni}= 0.059 \pm 0.001$\,M$_{\odot}$
and 
$t_{1}=312.9 \pm 4.6$\,days. The derived Ni mass is consistent with that calculated by~\cite{Singh_2023ixf}.
The estimated Ni mass of SN\,2023ixf is higher than that of SN\,1999em, which is 0.02\,M$_{\odot}$~\citep{Elmhamdi_1999em}.

\subsection{Comparison of Nebular Spectra with Theoretical Models} ~\label{sec:model comparison}
While the SN expands, its deep interior becomes visible as a result of decreased column densities and removal of free elections owing to recombination. 
Spectra at the nebular phase are hence free from photospheric absorption lines and are typically dominated by emission features. 
Nebular-phase spectra of CCSNe thus reveal the physical conditions of a variety of nuclear burning zones, enabling unambiguous model diagnostics.

\begin{figure}[h]
    \centering
    \includegraphics[width=\columnwidth]{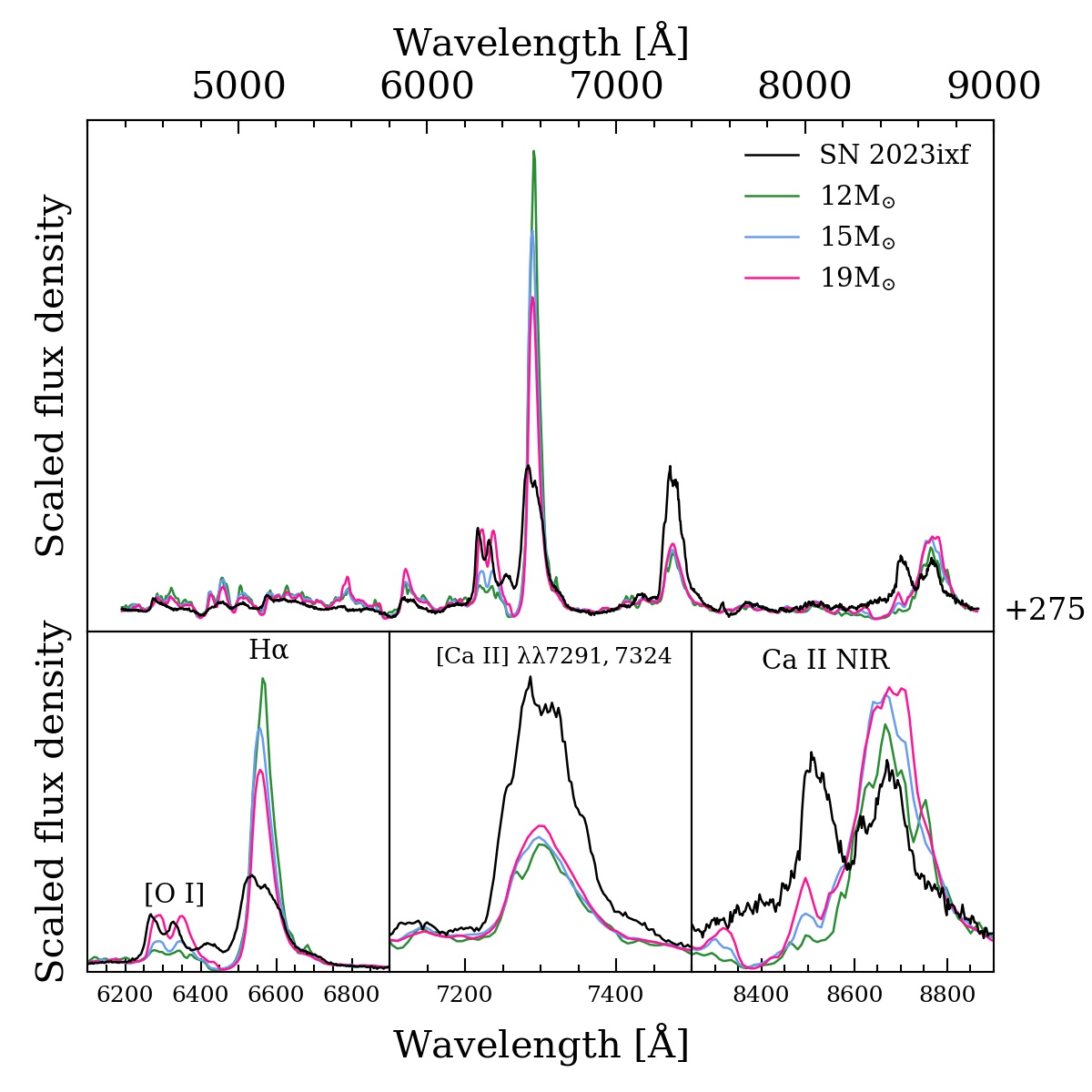}
    \caption{
    Comparison between the nebular-phase spectrum of SN\,2023ixf obtained at day 275 and the NLTE model spectra at day 250 for  CCSNe with ZAMS mass of 12, 15, and 19\,M$_{\odot}$. The subpanels in the lower row zoom in on the [O~{\sc i}] + H$\alpha$, [Ca~{\sc ii}], and Ca~{\sc ii} NIR triplet regions.
    }
    \label{fig:Jerkstrand}
\end{figure}

In Figure~\ref{fig:Jerkstrand}, we compare the $t \approx 275$\,day spectrum of SN\,2023ixf with a set of theoretical models with ZAMS masses of 12, 15, and 19\,M$_{\odot}$ at similar phases (at $t \approx 250$\,day;  \citealt{2014MNRAS.439.3694J}). The spectrum of SN\,2023ixf is flux calibrated using interpolated photometry.
For comparison purposes, 
all spectra have been scaled with respect to the total flux integrated over the observed wavelength range. We note that the profile of the [O~{\sc i}] $\lambda\lambda$6300, 6364 doublet of SN\,2023ixf is in good agreement with that calculated for the 15\,M$_{\odot}$ to 19\,M$_{\odot}$ models.
However, the [Ca~{\sc ii}] $\lambda\lambda$7291, 7323 doublet of SN\,2023ixf is a little bit stronger than that of the models, and H$\alpha$ of SN\,2023ixf is much weaker than the modeled profiles.
Note that the H$\alpha$, [Ca~{\sc ii}] $\lambda\lambda$7291, 7323, and Ca~{\sc ii} NIR triplet are mainly formed in the outer layers of the H-rich ejecta.
Consequently, their line strengths are dependent on both the pre-explosion mass loss and the distribution of the associated elements themselves. The latter may indicate the degree of outward mixing of Ni-rich ejecta.

\subsection{Progenitor Mass Estimation}~\label{sec:progenitor mass}
In particular, being one of the products of a series of hydrostatic burning phases, the amount of oxygen in the ejecta increases with ZAMS mass. Since oxygen also acts as the fuel for calcium production during the explosive burning phase, the flux ratio of [O~{\sc i}] $\lambda\lambda$6300, 6364 
to [Ca~{\sc ii}]/[O~{\sc i}]  
provides a sensitive indicator of the ZAMS mass of the progenitor of CCSNe~\citep{Fransson&Chevalier_1989, Jerkstrand_2012, 2014MNRAS.439.3694J, 2015A&A...573A..12J}. Our spectra of SN\,2023ixf have been scaled to match the photometry at different bandpasses. The photometry is interpolated or extrapolated to get the corresponding values at the phases the spectra were taken. The spectrum at day 298 is constructed from the spectra taken on days 297.2 and 298.1.

To derive the progenitor mass of SN\,2023ixf,  we first estimate the continuum underlying the [O~{\sc i}] line by connecting the visually inspected blue end of the [O~{\sc i}] and the red end of H$\alpha$. This ``pseudocontinuum'' represented by such a line segment is then  subtracted from the spectral region of interest, as shown in Fig.~\ref{fig:[O I]}.
Second, a multicomponent Gaussian function is used to fit the features of the [O~{\sc i}] + H$\alpha$ emission complex between days 275 and 407.
For the measurement of each phase, we calculate the area below the Gaussian functions that fit the $\lambda\lambda$6300, 6364 feature to obtain the [O~{\sc i}] luminosity. Finally, we divide the time series of [O~{\sc i}] luminosity by the intensity integrated in the range 4800--8900~\AA\ of the same spectrum.
The aim of the last step of normalization is to remove the dependence of [O~{\sc i}] luminosity on the $\gamma$-ray trapping. The selected wavelength range is similar to that adopted by \cite{2018NatAs...2..574A, Fang_ixf}.

\begin{figure}[h]
    \centering
    \includegraphics[width=\columnwidth]{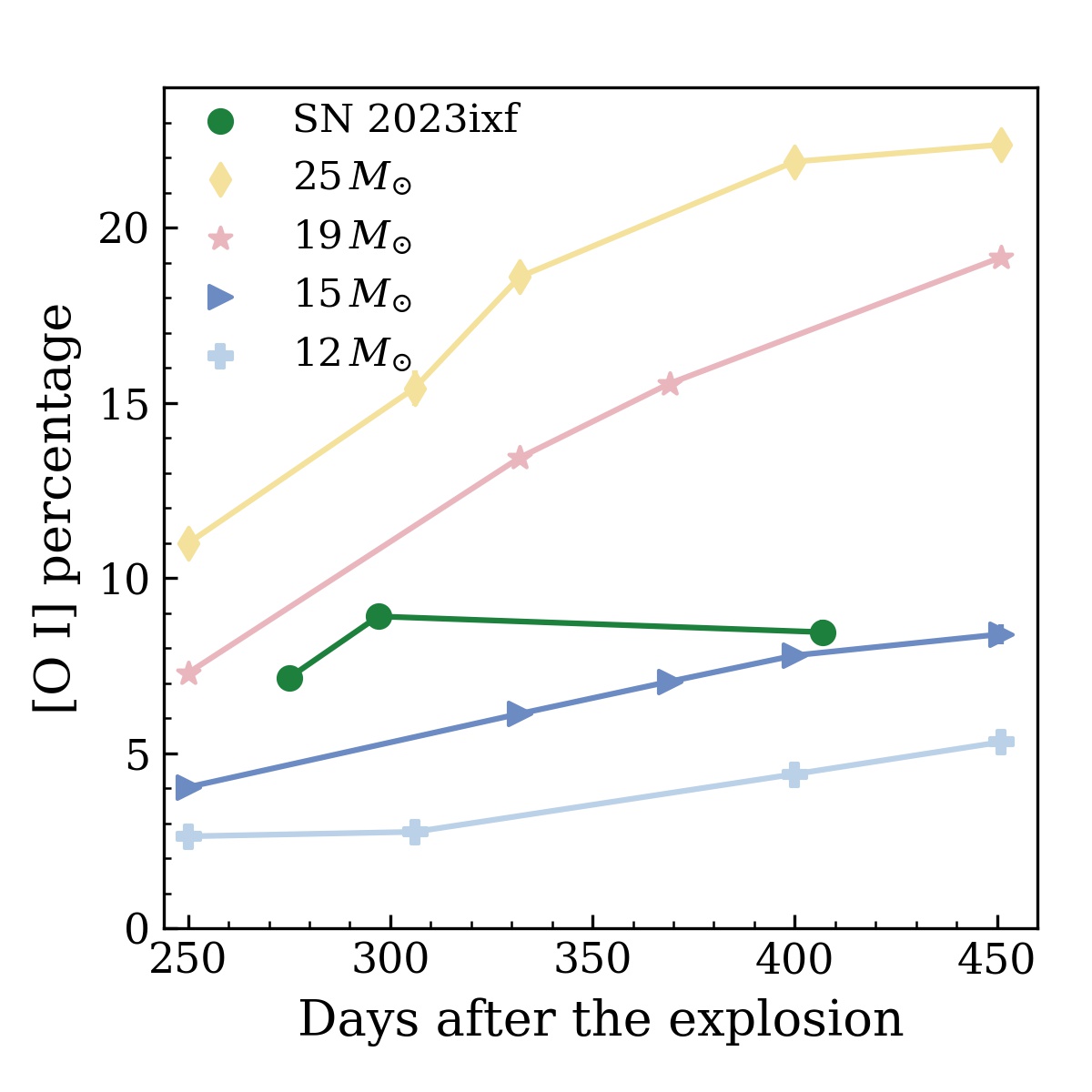}
    \caption{The temporal evolution of the measured [O~{\sc i}] $\lambda\lambda$6300, 6364 flux for SN\,2023ixf as a percentage of the total optical flux, along with that of the modeled spectra presented by \cite{2014MNRAS.439.3694J}.}
    \label{fig:progenitor_mass}
\end{figure}

In Figure~\ref{fig:progenitor_mass}, we present the fraction of the [O~{\sc i}] flux of SN\,2023ixf at days 275, 298, and 407.
The same calculation was also carried out for a series of model spectra with different progenitor ZAMS masses as computed by~\cite{2014MNRAS.439.3694J}.
The [O~{\sc i}] percentages of SN\,2023ixf 
estimated in the nebular phase
fall in the progenitor mass range of 15--19 M$_{\odot}$, consistent with the results of \cite{Fang_ixf, Jencson_progenitor, Niu_progenitor, Qin_progenitor, Zheng_23ixf}. 
 
We remark that the [O~{\sc i}] flux estimated based on multicomponent Gaussian fitting may introduce additional systematic uncertainties that are difficult to characterize. For example, any departure from spherical symmetry would alter the shape of the line profile.
Additionally, dust formation may alter the line asymmetry by suppressing the flux in the blue wing and creating an extended profile in the red wing.
A box-shaped underlying continuum of H$\alpha$ blended with [O~{\sc i}] may affect the estimation of the flux of [O~{\sc i}].
However, as the measurements of the data and the models were carried out following an identical procedure, and the intensity-normalized [O~{\sc i}] flux measured for various progenitor models confirms the trend of increasing mass of the progenitor, we suggest that the ZAMS mass estimated for SN\,2023ixf based on the [O~{\sc i}] flux remains robust.

\section{Conclusion}~\label{sec:conclusion}
We present extensive optical and NIR photometric and spectroscopic observations of SN\,2023ixf spanning 3--625 days after the explosion. The narrow flash lines of SN\,2023ixf persist for the first eight days. These long-lived features indicate a radially extended CSM shell around the progenitor. SN\,2023ixf reaches a $V$-band absolute peak magnitude of $M_V=-18.1\pm0.1$\,mag about 6 days after the explosion, which is toward the bright end of Type II SNe.
SN\,2023ixf presents a slanted plateau phase ($\sim$1.47\,mag per 100\,d in $V$), indicating a less massive H envelope than those of normal Type IIP SNe. 

The photospheric velocities of SN\,2023ixf are higher than those of typical SNe IIP, suggesting higher ejecta kinetic energy. The metallicity of SN\,2023ixf derived from the the pEW of Fe~{\sc ii} during the photospheric phase falls between 0.4\,Z$_\odot$ and 1\,Z$_\odot$. Using EPM, we estimate the distance to the SN as $6.35^{+0.31}_{-0.39}$\,Mpc.

Taking into account the incomplete trapping of the $\gamma$-rays, we use the radioactive-tail luminosity of SN\,2023ixf to calculate the mass of $^{56}$Ni synthesized during the SN explosion, 
The derived mass is $0.059 \pm 0.001$\,M$_{\odot}$.
We compare the percentage of [O~{\sc i}] flux with that of the synthetic spectra in~\cite{2014MNRAS.439.3694J}; the derived progenitor mass of SN\,2023ixf is 15--19\,M$_{\odot}$, consistent with the results suggested from {\it HST} and {\it Spitzer} archival images. The emission lines of hydrogen, sodium, oxygen, and calcium exhibit double-peaked profiles during the nebular phase, hinting at an aspherical distribution of the ejecta. The box-shaped H$\alpha$ profile extending up to 8000\,km\,s$^{-1}$ indicates  interaction between the ejecta and CSM. The different profiles on the blue and red sides of the H$\alpha$ profile may be attributed to a spherically asymmetric CSM shell. The disappearance of the red shoulder of H$\alpha$ at day $\sim$150 and its reappearance at day $\sim$275 might result from multiple episodes of mass loss $\gtrsim$200\,yr before the explosion.

\section{Acknowledgments}
 This work is supported by the National Science Foundation of China (NSFC grants 12288102 and 12033003) and the Tencent Xplorer Prize. A.R. acknowledges financial support from the GRAWITA Large Program Grant (PI P. D'Avanzo) and from PRIN-INAF 2022 "Shedding light on the nature of gap transients: from the observations to the models."
Y.-Z. Cai is supported by NSFC grant 12303054, the National Key Research and Development Program of China (grant 2024YFA1611603), the Yunnan Fundamental Research Projects (grant 202401AU070063), and the International Centre of Supernovae, Yunnan Key Laboratory (grant 202302AN360001). A.P., A.R., G.V., and I.S. are supported by PRIN-INAF 2022 project "Shedding light on the nature of gap transients: from the observations to the models."
J.Z. is supported by the National Key R\&D Program of China with grant 2021YFA1600404, NSFC grants 12173082 and 12333008, the Yunnan Fundamental Research Projects (grants 202501AV070012, 202401BC070007 and 202201AT070069), the Top-notch Young Talents Program of Yunnan Province, the Light of West China Program provided by the Chinese Academy of Sciences, and the International Centre of Supernovae, Yunnan Key Laboratory (grant 202302AN360001). M.H. acknowledges the support from the Postdoctoral Fellowship Program of CPSF under Grant Number GZB20240376, and the Shuimu Tsinghua Scholar
Program 2024SM118.
A.V.F.'s group at UC Berkeley is grateful for financial support from the Christopher R. Redlich Fund, Gary and Cynthia Bengier, Clark and Sharon Winslow, Sanford Robertson (Y.Y. was a Bengier-Winslow-Robertson Postdoctoral Fellow when this work was started), and many other donors.

We acknowledge the support of the staff of the Xinglong 80cm telescope and the Xinglong 2.16m telecope. This work was partially supported by the Open Project Program of the Key Laboratory of Optical Astronomy, National Astronomical Observatories, Chinese Academy of Sciences.

Some of the data presented herein were obtained at the W. M. Keck Observatory, which is operated as a scientific partnership among the California Institute of Technology, the University of California, and NASA; the observatory was made possible by the generous financial support of the W. M. Keck Foundation.
We thank Melissa Graham, Patrick Kelly, WeiKang Zheng, and Jon Mauerhan for assistance obtaining the Keck SN\,2013ej spectrum.

\newpage
\bibliography{reference}{}
\bibliographystyle{aasjournal}

\begin{appendix}
\onecolumn
\section{Spectroscopic observations}
\begin{longtable}{cccccc}
\caption{Log of optical spectroscopic observations of SN\,2023ixf}\label{table:Log of optical spectroscopy}\\

\hline
UTC time & MJD & Phase & Range & Resolution & Telescope/Instrument \\ 
 (yyyy-mm-dd hh:mm) & & (d) & (\AA)&(blue/red)(\AA) & \\
\hline
\endfirsthead
\multicolumn{6}{c}%
{{\bfseries \tablename\ \thetable{} -- Continued}} \\
\hline 
UT time & MJD & Phase & Range & Resolution & Telescope/Instrument \\ 
 (yyyy-mm-dd hh:mm) & & (d) & (\AA)&(blue/red)(\AA) & \\
\hline
\endhead
\hline 
\multicolumn{6}{r}{{Continued}} \\ \hline
\endfoot
\hline
\endlastfoot
2023-05-21 13:54 & 60085.58 & 2.79 & 3895-8871 & 4.45 & XLT/BFOSC  \\ 
2023-05-26 12:49 & 60090.53 & 7.75 & 3894-8870 & 4.45 & XLT/BFOSC  \\ 
2023-05-31 12:39 & 60095.53 & 12.74 & 3892-8869 & 4.45 & XLT/BFOSC  \\ 
2023-06-01 13:01 & 60096.54 & 13.75 & 3889-8867 & 4.45 & XLT/BFOSC  \\ 
2023-06-07 20:31 & 60102.86 & 20.07 & 4250-6590 & 3.6 & GT/B\&C  \\ 
2023-06-08 21:00 & 60103.88 & 21.09 & 3320-7890 & 5.8 & GT/B\&C  \\ 
2023-06-09 00:45 & 60104.03 & 21.24 & 3320-7890 & 5.6 & GT/B\&C  \\ 
2023-06-12 02:12 & 60107.09 & 24.30 & 3660-6060 & 4.1 & GT/B\&C  \\ 
2023-06-14 01:34 & 60109.07 & 26.28 & 3320-7890 & 5.6 & GT/B\&C  \\ 
2023-06-15 23:45 & 60110.99 & 28.20 & 3320-7890 & 5.6 & GT/B\&C  \\ 
2023-06-16 22:17 & 60111.93 & 29.14 & 3320-7890 & 5.7 & GT/B\&C  \\ 
2023-06-17 21:01 & 60112.88 & 30.09 & 3320-7890 & 5.5 & GT/B\&C  \\ 
2023-06-18 20:43 & 60113.83 & 31.05 & 3320-7890 & 5.6 & GT/B\&C  \\ 
2023-06-21 16:56 & 60116.71 & 33.92 & 6004-8264 & 1.79 & XLT/BFOSC  \\ 
2023-06-21 17:29 & 60116.73 & 33.94 & 3900-8875 & 4.45 & XLT/BFOSC  \\ 
2023-06-22 21:37 & 60118.90 & 36.11 & 4250-6590 & 3.4 & GT/B\&C  \\ 
2023-06-24 13:09 & 60119.55 & 36.76 & 5995-8255 & 1.79 & XLT/BFOSC  \\ 
2023-06-24 13:40 & 60119.57 & 36.78 & 3913-8890 & 4.45 & XLT/BFOSC  \\ 
2023-06-25 20:36 & 60120.86 & 38.07 & 3320-7890 & 5.5 & GT/B\&C  \\ 
2023-06-27 00:45 & 60122.03 & 39.24 & 4460-6860 & 3.7 & GT/B\&C  \\ 
2023-06-27 13:45 & 60122.57 & 39.79 & 3936-8921 & 4.45 & XLT/BFOSC  \\ 
2023-06-27 14:05 & 60122.59 & 39.80 & 3427-5470 & 1.98 & XLT/BFOSC  \\ 
2023-06-27 14:36 & 60122.61 & 39.82 & 6006-8268 & 1.79 & XLT/BFOSC  \\ 
2023-06-29 14:13 & 60124.59 & 41.80 & 3905-8885 & 4.45 & XLT/BFOSC  \\ 
2023-06-29 22:07 & 60124.92 & 42.13 & 3320-7890 & 5.5 & GT/B\&C  \\ 
2023-06-30 14:42 & 60125.61 & 42.83 & 6006-8267 & 1.79 & XLT/BFOSC  \\ 
2023-07-01 20:59 & 60126.88 & 44.09 & 5775-6975 & 1.3 & GT/B\&C  \\ 
2023-07-10 22:50 & 60135.95 & 53.16 & 3670-6070 & 4.1 & GT/B\&C  \\ 
2023-07-17 20:11 & 60142.84 & 60.05 & 3310-7890 & 6.9 & GT/B\&C  \\ 
2023-07-18 23:33 & 60143.98 & 61.19 & 3310-7890 & 7.3 & GT/B\&C  \\ 
2023-07-23 20:11 & 60148.84 & 66.05 & 3310-7890 & 6.7 & GT/B\&C  \\ 
2023-07-26 14:53 & 60151.62 & 68.83 & 3800-8750 & 2.85 & LJT/YFOSC  \\ 
2023-08-06 20:02 & 60162.83 & 80.05 & 3310-7890 & 7.1 & GT/B\&C  \\ 
2023-08-12 20:07 & 60168.84 & 86.05 & 3310-7890 & 6.9 & GT/B\&C  \\ 
2023-08-13 19:54 & 60169.83 & 87.04 & 3310-7890 & 6.6 & GT/B\&C  \\ 
2023-08-18 21:22 & 60174.91 & 92.12 & 3320-7890 & 6.9 & GT/B\&C  \\ 
2023-08-19 21:02 & 60175.88 & 93.09 & 4000-9300 & 14.7 & Copernico/AFOSC  \\ 
2023-08-19 22:10 & 60175.92 & 93.14 & 3320-7890 & 7.0 & GT/B\&C  \\ 
2023-08-20 21:31 & 60176.90 & 94.11 & 3320-7890 & 6.9 & GT/B\&C  \\ 
2023-08-21 20:50 & 60177.87 & 95.08 & 3320-7890 & 7.2 & GT/B\&C  \\ 
2023-08-24 20:25 & 60180.85 & 98.06 & 4250-6590 & 3.3 & GT/B\&C  \\ 
2023-08-31 19:08 & 60187.80 & 105.01 & 3320-7890 & 7.5 & GT/B\&C  \\ 
2023-09-01 20:51 & 60188.87 & 106.08 & 3320-7890 & 6.8 & GT/B\&C  \\ 
2023-09-06 19:14 & 60193.80 & 111.01 & 3320-7890 & 7.4 & GT/B\&C  \\ 
2023-09-08 20:00 & 60195.83 & 113.05 & 5775-6975 & 1.5 & GT/B\&C  \\ 
2023-09-18 11:11 & 60205.47 & 122.68 & 3899-8876 & 4.45 & XLT/BFOSC  \\ 
2023-09-24 19:08 & 60211.80 & 129.01 & 3320-7890 & 7.1 & GT/B\&C  \\ 
2023-09-27 11:37 & 60214.48 & 131.70 & 3939-8918 & 4.45 & XLT/BFOSC  \\ 
2023-09-28 20:05 & 60215.84 & 133.05 & 3320-7890 & 6.2 & GT/B\&C  \\ 
2023-09-29 19:11 & 60216.80 & 134.01 & 3320-7890 & 6.1 & GT/B\&C  \\ 
2023-10-06 18:19 & 60223.76 & 140.98 & 4580-7000 & 3.4 & GT/B\&C  \\ 
2023-10-09 10:53 & 60226.45 & 143.67 & 3939-8921 & 4.45 & XLT/BFOSC  \\ 
2023-10-09 18:00 & 60226.75 & 143.96 & 3320-7890 & 7.3 & GT/B\&C  \\ 
2023-10-11 10:38 & 60228.44 & 145.66 & 6006-8275 & 1.79 & XLT/BFOSC  \\ 
2023-10-11 18:05 & 60228.75 & 145.97 & 3320-7890 & 7.2 & GT/B\&C  \\ 
2023-10-16 18:11 & 60233.76 & 150.97 & 3530-9290 & 21.6/23.1 & Copernico/AFOSC  \\ 
2023-10-19 10:27 & 60236.44 & 153.65 & 3943-8922 & 4.45 & XLT/BFOSC  \\ 
2023-10-22 17:27 & 60239.73 & 156.94 & 3320-7890 & 6.4 & GT/B\&C  \\ 
2023-10-25 04:38 & 60242.19 & 159.41 & 3320-7890 & 5.7 & GT/B\&C  \\ 
2023-10-28 17:45 & 60245.74 & 162.95 & 3330-7890 & 6.5 & GT/B\&C  \\ 
2023-11-05 17:11 & 60253.72 & 170.93 & 3320-7890 & 6.65 & GT/B\&C  \\ 
2023-11-11 04:43 & 60259.20 & 176.41 & 3320-7890 & 5.7 & GT/B\&C  \\ 
2023-11-19 04:16 & 60267.18 & 184.39 & 3250-9280 & 15.4/14.5 & Copernico/AFOSC  \\ 
2023-11-22 04:26 & 60270.18 & 187.40 & 5775-6975 & 2.1 & GT/B\&C  \\ 
2023-11-23 04:19 & 60271.18 & 188.39 & 3320-7890 & 7.4 & GT/B\&C  \\ 
2023-11-29 03:30 & 60277.15 & 194.36 & 5775-6975 & 2.5 & GT/B\&C  \\ 
2023-12-02 21:13 & 60280.88 & 198.10 & 3775-8918 & 4.45 & XLT/BFOSC  \\ 
2023-12-03 03:28 & 60281.14 & 198.36 & 3315-7888 & 7.5 & GT/B\&C  \\ 
2023-12-07 02:11 & 60285.09 & 202.30 & 4980-9280 & 23 & Copernico/AFOSC  \\ 
2023-12-16 03:53 & 60294.16 & 211.37 & 5775-6975 & 2.4 & GT/B\&C  \\ 
2023-12-17 01:33 & 60295.07 & 212.28 & 4980-9280 & 15 & Copernico/AFOSC  \\ 
2023-12-18 21:10 & 60296.88 & 214.09 & 3778-8919 & 4.45 & XLT/BFOSC  \\ 
2023-12-21 04:56 & 60299.21 & 216.42 & 4890-7305 & 5.4 & GT/B\&C  \\ 
2023-12-31 20:54 & 60309.87 & 227.08 & 3774-8913 & 4.45 & XLT/BFOSC  \\ 
2024-01-04 04:40 & 60313.19 & 230.41 & 5775-6975 & 2.7 & GT/B\&C  \\ 
2024-01-13 03:21 & 60322.14 & 239.35 & 5775-6975 & 3.6 & GT/B\&C  \\ 
2024-01-16 02:10 & 60325.09 & 242.30 & 3320-7890 & 6.5 & GT/B\&C  \\ 
2024-02-17 21:06 & 60357.88 & 275.09 & 3800-8750 & 2.85 & LJT/YFOSC  \\ 
2024-03-10 22:59 & 60379.96 & 297.17 & 4000-9300 & 14.7 & Copernico/AFOSC  \\ 
2024-03-11 21:30 & 60380.90 & 298.11 & 3600-7300 & 24.5 & Copernico/AFOSC \\
2024-04-07 01:00 & 60407.04 & 324.25 & 4000-9300 & 14.7 & Copernico/AFOSC \\
\hline
\end{longtable}

\begin{longtable}{cccccc}
\caption{Log of NIR spectroscopic observations of SN\,2023ixf}\label{table:Log of NIR spectroscopy}\\
\hline
UTC time & MJD & Phase & Range & Resolution & Telescope/Instrument \\ 
 (yyyy-mm-dd hh:mm) & & (d) & (\AA)&(blue/red)(\AA) & \\
\hline
\endfirsthead
\multicolumn{6}{c}%
{{\bfseries \tablename\ \thetable{} -- Continued}} \\
\hline 
UT time (yyyy-mm-dd) & MJD & Phase(d) & Range & Resolution & Telescope/Instrument\\
\hline
\endhead
\hline 
\multicolumn{6}{r}{{Continued}} \\ \hline
\endfoot
\hline
\endlastfoot
2023-05-23 01:50 & 60087.08 & 4.29 & 10850-24776 & 6.6/11.2 & TNG/NICS \\
2023-05-29 21:16 & 60093.89 & 11.10 & 10807-24743 & 6.6/11.2 & TNG/NICS \\
2023-06-02 23:04 & 60097.96 & 15.17 & 10830-24721 & 6.6/11.2 & TNG/NICS \\
2023-06-20 23:09 & 60115.97 & 33.18 & 10783-24735 & 6.6/11.2 & TNG/NICS \\
2023-07-06 22:58 & 60131.96 & 49.17 & 10781-24735 & 6.6/11.2 & TNG/NICS \\
2023-07-27 21:33 & 60152.90 & 70.11 & 10784-24743 & 6.6/11.2 & TNG/NICS \\
2023-08-07 20:55 & 60163.87 & 81.08 & 10788-24741 & 6.6/11.2 & TNG/NICS \\
\hline
\end{longtable}
\newpage
\begin{figure*}[h]
    \centering
    \caption{Spectral time series of SN\,2023ixf spanning days 3 to 324. Observations conducted with different instruments are color coded and distinguished by the legend at the top. Phases are marked on the right. 
    Some major telluric lines are identified by $\oplus$. Some noisy spectra are binned, with the original version plotted with a fainter color.}
    \includegraphics[width=0.9\textwidth]{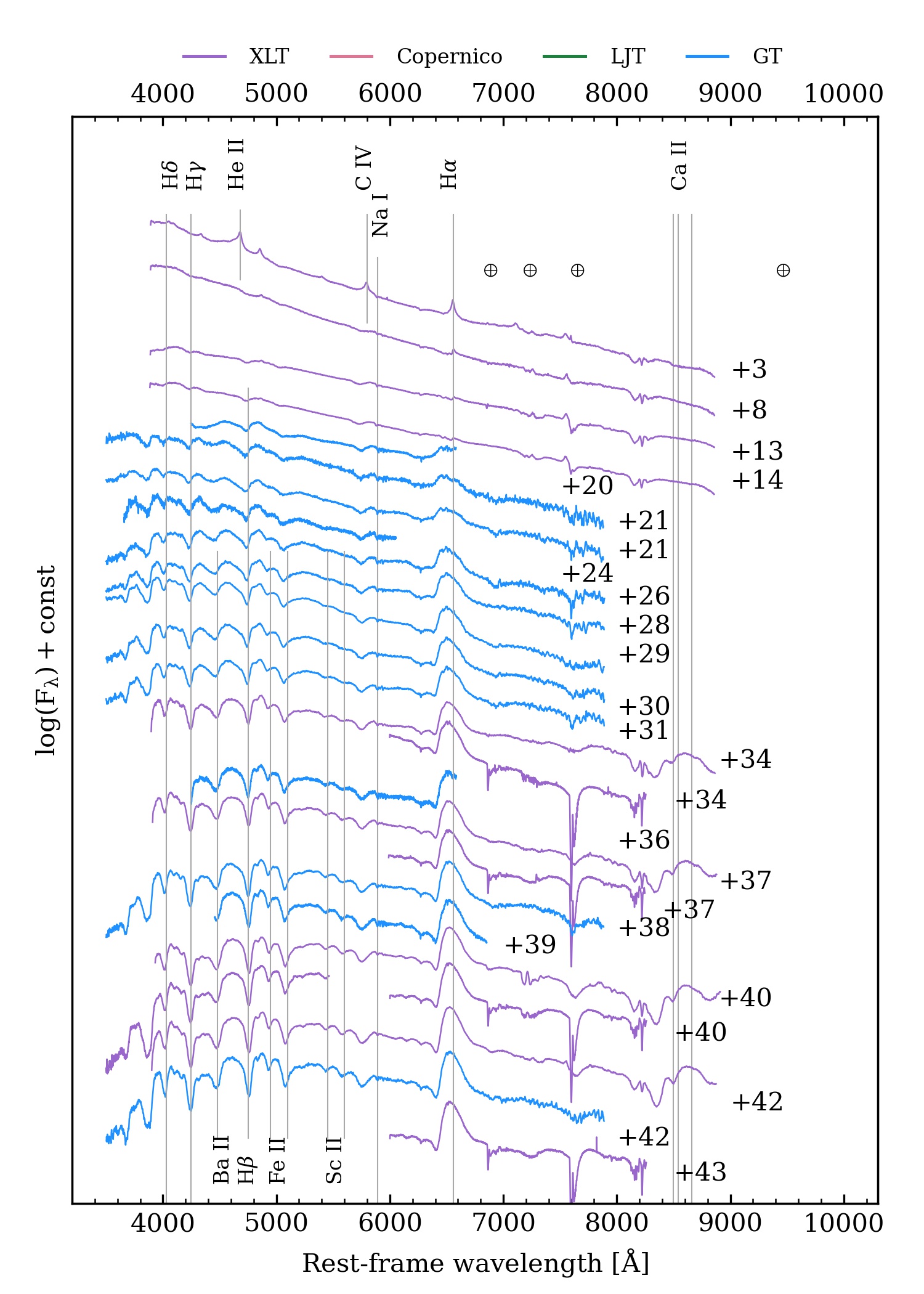}
    \newpage 
    \begin{center} 
    Continued
    \end{center}

    \label{fig:spectra}
\end{figure*}

\begin{figure*}[h]
    \centering
    \includegraphics[width=0.9\textwidth]{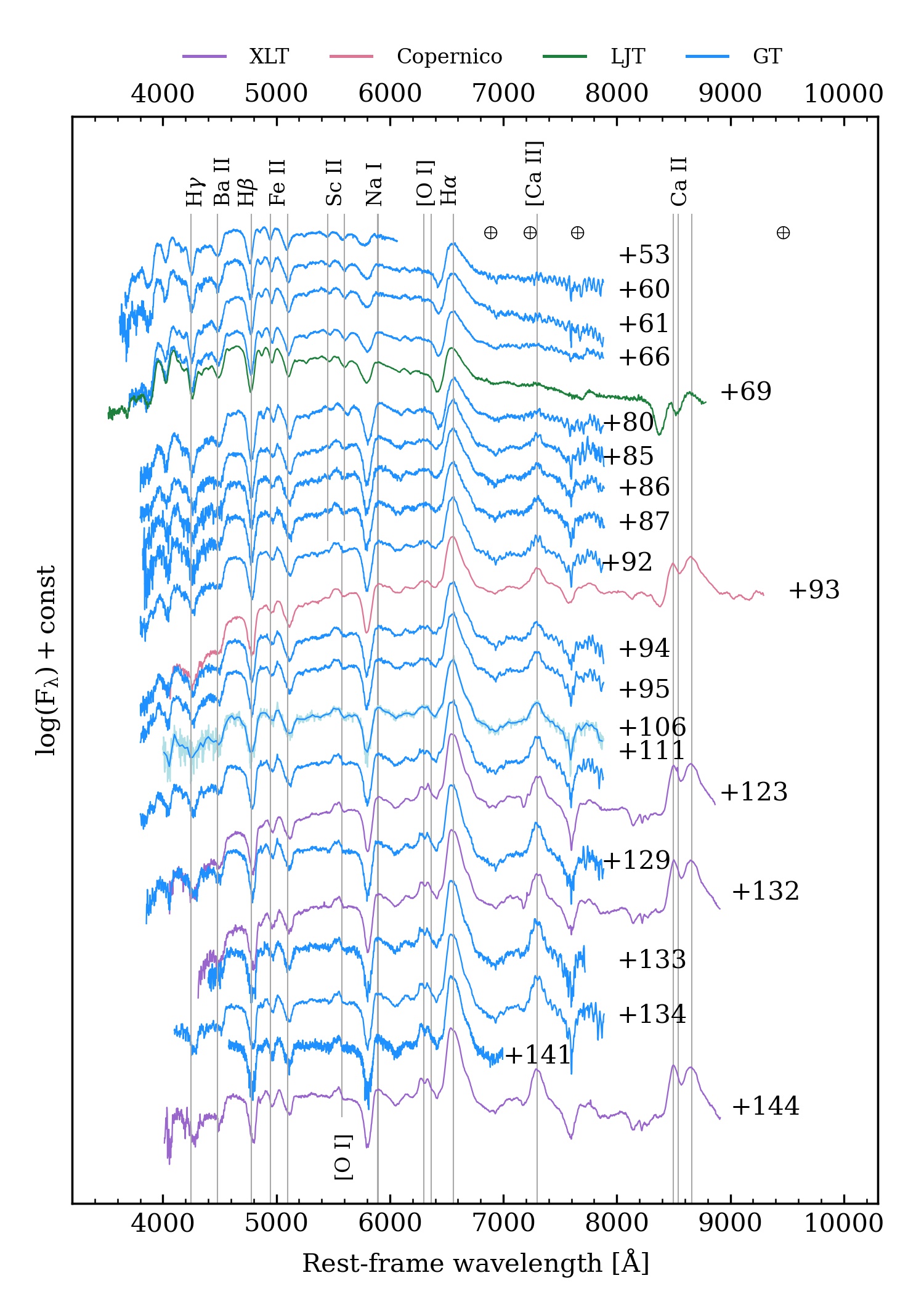}
    \newpage
    \begin{center} 
    Continued
    \end{center}
\end{figure*}

\begin{figure*}[h]
    \centering
    \includegraphics[width=0.9\textwidth]{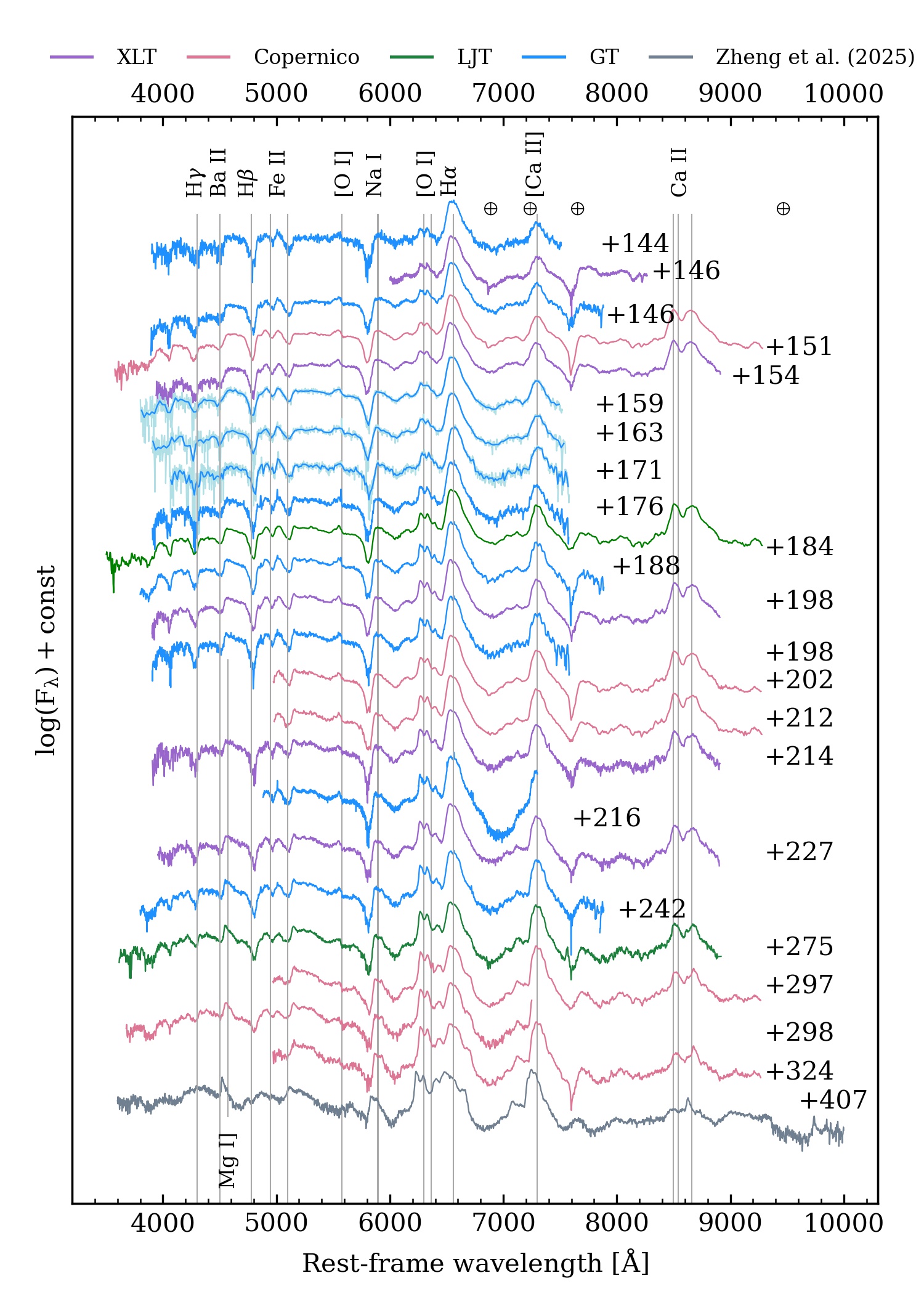}
    \newpage
\end{figure*}
\clearpage

\section{Fitting [O~{\sc i}] + H$\alpha$ with Multiple Gaussians}
\begin{figure*}[h]
    \centering
    \includegraphics[width=0.9\textwidth]{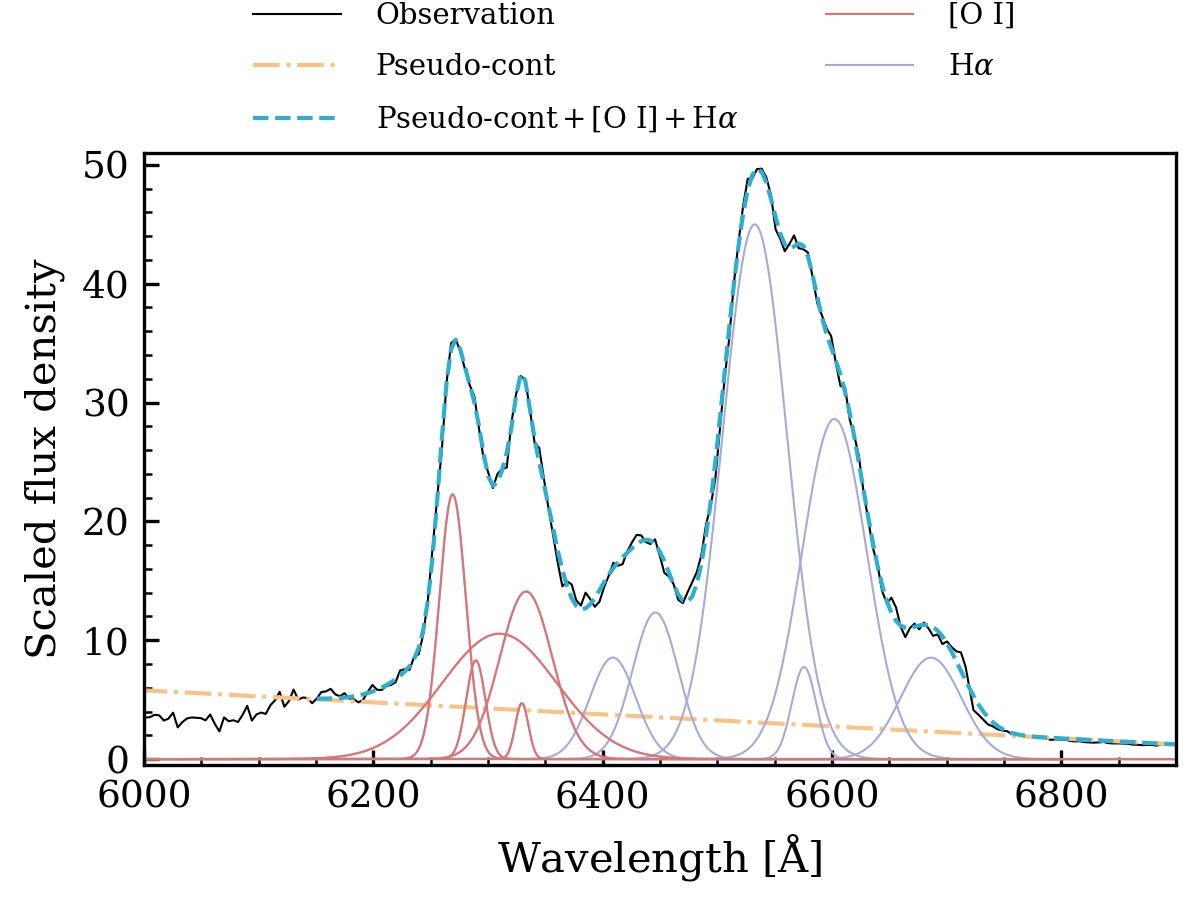}
    \caption{The [O~{\sc I}] + H$\alpha$ profile of SN\,2023ixf at day 298. The observed spectrum (black solid line) is approximated by the dashed light-blue line, which consists of a pseudocontinuum (orange dot-dashed line), five Gaussian components of [O~{\sc I}] (solid red lines), and five Gaussian components of H$\alpha$ (solid dark-blue lines).}
    \label{fig:[O I]}
\end{figure*}

\end{appendix}

\end{document}